\documentclass[journal]{IEEEtran}

\usepackage{amsmath,amstext,amsfonts,amssymb,eucal,graphicx}
\usepackage{subfigure}
\usepackage{epsfig}

\newtheorem{lemma}{Lemma}

\newtheorem{proposition}{Proposition}
\newtheorem{theorem}{Theorem}
\newtheorem{definition}{Definition}

\newcommand{\sluttlinje}{\hspace*{1cm}\hspace*{-1cm}~\hfill}
\newcommand{\sluttmerke}{\sluttlinje\raisebox{-1mm}{\rule{2.5mm}{2.5mm}}}

\newcommand{\boxslash}{\begin{picture}(9,5)  \put(2,0){\framebox(5,5){$\smallsetminus$}} \end{picture}}

\newcommand{\N}{\mathbb{N}}

\begin{document}
\bibliographystyle{IEEEtran}
\title{Channel Capacity Estimation using Free Probability Theory}

\author{
  \IEEEauthorblockN{\O yvind Ryan,~\IEEEmembership{Member,~IEEE} and M{\'e}rouane~Debbah,~\IEEEmembership{Member,~IEEE}\\}
  \thanks{This work was supported by Alcatel-Lucent within the Alcatel-Lucent Chair on flexible radio at SUPELEC}
  \thanks{This paper was presented in part at the Asilomar Conference on Signals, Systems and Computers, 2007, Pacific Grove, USA}
  \thanks{\O yvind~Ryan is with the Centre of Mathematics for Applications, University of Oslo, P.O. Box 1053 Blindern, NO-0316 Oslo, NORWAY, oyvindry@ifi.uio.no}
  \thanks{M{\'e}rouane~Debbah is with SUPELEC, Gif-sur-Yvette, France, merouane.debbah@supelec.fr}
}
\maketitle

\markboth{IEEE Transactions on Signal Processing,~Vol.~1,
No.~1,~January~2007}{Ryan \MakeLowercase{\textit{et al.}}: Channel Capacity Estimation using Free Probability Theory}

\maketitle
\begin{abstract}
In many channel measurement applications, one needs to estimate some
characteristics of the channels based on a limited set of
measurements. This is mainly due to the highly time varying
characteristics of the channel. In this contribution, it will be
shown how free probability can be used for channel capacity
estimation in MIMO systems. Free probability has already been
applied in various application fields such as digital
communications, nuclear physics and mathematical finance, and has
been shown to be an invaluable tool for describing the asymptotic
behaviour of many large-dimensional systems. In particular, using
the concept of free deconvolution, we provide an asymptotically
(w.r.t. the number of observations) unbiased capacity estimator
 for MIMO channels impaired with noise called the free probability based estimator.
Another estimator, called the Gaussian matrix mean based estimator, is also introduced
by slightly modifying the free probability based estimator.
This estimator is shown to give unbiased estimation of the moments of the channel matrix
for any number of observations. Also, the estimator has this property when we extend to MIMO
channels with phase off-set and frequency drift, for which no
estimator has been provided so far in the literature.
It is also shown that both the free probability based and the Gaussian matrix mean based estimator
are asymptotically unbiased capacity estimators as the number of transmit antennas go to infinity,
regardless of whether phase off-set and frequency drift are present.
The limitations in the two estimators are also explained.
Simulations are run to assess the performance of the estimators for a low number of
antennas and samples to confirm the usefulness of the asymptotic results.
\end{abstract}


\section{Introduction}
Random matrices, and in particular limit distributions of sample
covariance matrices, have proved to be a useful tool for modelling
systems, for instance in digital communications
\cite{paper:telatar99}, nuclear physics~\cite{paper:guhr} and
mathematical finance~\cite{book:bouchaud}. A typical random matrix
model is the information-plus-noise model,
\begin{equation} \label{system1}
  {\bf W}_n = \frac{1}{N}({\bf R}_n + \sigma {\bf X}_n)({\bf R}_n + \sigma {\bf X}_n)^{H}.
\end{equation}
${\bf R}_n$ and ${\bf X}_n$ are assumed independent random matrices
of dimension $n\times N$, where ${\bf X}_n$ contains i.i.d. standard
(i.e. mean $0$, variance $1$) complex Gaussian entries.
(\ref{system1}) can be thought of as the sample covariance matrices
of random vectors ${\bf r}_n + \sigma {\bf x}_n$. ${\bf r}_n$ can be
interpreted as a vector containing the system characteristics
(direction of arrival for instance in radar applications or impulse
response in channel estimation applications). ${\bf x}_n$ represents
additive noise, with $\sigma$ a measure of the strength of the
noise. Classical signal processing estimation methods consider the
case where the number of observations $N$ is highly bigger than the
dimensions of the system $n$, for which equation (\ref{system1})
can be shown to be approximately:
\begin{equation} \label{system2}
  {\bf W}_n = {\bf \Gamma}_n +\sigma^2 {\bf I}_n.
\end{equation}
Here, ${\bf \Gamma}_n$ is the true covariance of the signal. In this
case, one can separate the signal eigenvalues from the noise ones
and infer (based only on the statistics of the signal) on the
characteristics of the input signal.  However, in  many situations,
one can gather only a limited number of observations during which
the characteristics of the signal does not change. In order to model
this case, $n$ and $N$ will be increased so that
$\lim_{n\rightarrow\infty}\frac{n}{N} = c$, i.e. the number of
observations is increased at the same rate as the number of
parameters of the system (note that equation (\ref{system2})
corresponds to the case $c=0$).

Previous contributions have already dealt with this problem.
In~\cite{paper:doziersilverstein1}, Dozier and Silverstein explain
how one can use the eigenvalue distribution of ${\bf
\Gamma}_n=\frac{1}{N}{\bf R}_n{\bf R}_n^{H}$ to estimate the
eigenvalue distribution of ${\bf W}_n$ by solving a given equation.
In~\cite{eurecom:multfreeconv,eurecom:freedeconvinftheory}, we
provided an algorithm for passing between the two, using the concept
of  {\em multiplicative free convolution}, which admits a convenient
implementation. The implementation performs free convolution exactly
based solely on moments.

In this paper, channel capacity estimation in MIMO systems is used
as a benchmark application by using the connection between free
probability theory and systems of type (\ref{system1}). For MIMO
channels with and without frequency off-sets, we derive explicit
asymptotically unbiased estimators which perform much better than classical ones.
We do not prove directly that the proposed estimators work better than the classical ones, 
but present simulations which indicate that they are superior. 
We remark that the proposed capacity estimators will not be unbiased, 
it is needed that either the number of transmit antennas or the number of observations be large to obtain precise estimation. 
This limitation is most severe for channels with frequency off-sets, where it is needed in any case that the number of transmit antennas is large 
to obtain precise estimation. 
A case of study where channel estimation using free deconvolution has been used can be found in~\cite{newfrommerouane1} and~\cite{newfrommerouane2}.

This paper is organized as follows.
Section~\ref{statementoftheproblem} presents the problem under
consideration. Section~\ref{framework} provides the basic concepts
needed on free probability, including free convolution. 
In section~\ref{newestimators}, we formalize a new channel capacity estimator 
based on free probability, and explain some of the
shortcomings for MIMO models with frequency off-sets. Another
estimator, called the Gaussian matrix mean based estimator is then formalized
to address the shortcomings of the free probability based estimator. 
We also present arguments for the Gaussian matrix mean based estimator performing better than the free probability based estimator, 
in some specific cases. These arguments are, however, not definite;
we do not prove that one estimator is better than the other for the cases considered. 
The limitations of the estimators are also explained. 
The low rank of the channel (less than or equal to four) is the most notable limitation. 
In section~\ref{simulations}, simulations of the 
estimators are performed and compared, where several quantities are varied, 
like the noise variance, rank and dimensions of the channel matrix, and the number
of observations. In the following, upper (lower boldface) symbols
will be used for matrices (column vectors) whereas lower symbols
will represent scalar values, $(.)^T$ will denote transpose
operator, $(.)^\star$ conjugation and
$(.)^H=\left((.)^T\right)^\star$ hermitian transpose. ${\bf I}_n$
will represent the $n\times n$ identity matrix. $Tr_n$ will denote
the non-normalized trace on $n\times n$ matrices, while $tr_n =
\frac{1}{n} Tr_n$ denotes the normalized trace. Also, we will
throughout the paper use $c$ as a shorthand notation for the ratio
between the number of rows and the number of columns in the random
matrix model being considered.

\section{Statement of the problem}
\label{statementoftheproblem} In usual time varying measurement
methods for MIMO systems, one validates models
\cite{ieeeJSAC.Kermoal02} by determining how the model fits with
actual capacity measurements. In this setting, one has to be
extremely cautious about the measurement noise, especially for far
field measurements  where the signal strength can be lower than the
noise.

The MIMO measured channel in the frequency domain can be modelled by~\cite{channelmodel1,channelmodel2}
\begin{eqnarray} \label{model2}
  \hat{{\bf H}}_i =  {\bf D}_i^{r}{\bf  H} {\bf D}_i^{t} + \sigma {\bf X}_i
\end{eqnarray}
where $\hat{\bf H}_i$,  ${\bf H}$ and  ${\bf X}_i$ are respectively
the $n \times m$ measured MIMO matrix ($n$  is the number of
receiving antennas, $m$ is the number of transmitting antennas), the
$n \times m$ MIMO channel and the $n \times m$ noise matrix with
i.i.d. standard Gaussian entries. Note that we suppose the noise
matrix ${\bf X}_i$ to be spatially white. In the realm of the
channel measurements under study, the antenna outputs are connected
to different RF (Radio Frequency) chains. As a consequence, for the
case under study, the channel noise impairments are independent from
one received antenna to the other. When one RF chain is used, the
noise to be considered is not white. This case can also be studied
within the framework of free deconvolution but goes beyond the scope
of the paper. We suppose that the channel ${\bf H}$, although time
varying, stays constant (block fading assumption) during $L$ blocks.
${\bf D}_i^{r}$ and ${\bf D}_i^{t}$ are $n\times n$ and $m\times m$
diagonal matrices which represent phase off-sets and phase drifts
(which are impairments due to the antennas and not the channel) at
the receiver and transmitter given respectively by (these are
supposed to vary on a block basis)
\begin{eqnarray*}
{\bf D}_i^{r} &=& \mathrm{diag}[e^{j\phi_1^i},...,e^{j\phi_n^i}] \mbox{, and} \\
{\bf D}_i^{t} &=& \mathrm{diag}[e^{j\theta_1^i},...,e^{j\theta_m^t}]
\end{eqnarray*}
where the phases $\phi_j^i$ and $\theta_j^i$ are random. We 
assume all phases independent and uniformly distributed.

We will also compare (\ref{model2}) with the simpler model
\begin{eqnarray} \label{model1}
  \hat{{\bf H}}_i = {\bf  H} + \sigma {\bf X}_i,
\end{eqnarray}
which is (\ref{model2}) without phase off-sets and phase drifts.

The capacity per receiving antenna (in the case where the noise is
spatially white additive Gaussian and the channel is not known at the
transmitter) of a channel with channel matrix ${\bf H}$ and signal
to noise ratio $\rho=\frac{1}{\sigma^2}$ is given by
\begin{equation} \label{channelcapacitydef}
  C 
  =
  \frac{1}{n}\log_2 \det\left( {\bf I}_n+\frac{1}{m \sigma^2} {\bf H}{\bf H}^{H}\right)
  = 
  \frac{1}{n} \sum_{l=1}^{n} \log_2(1+ \frac{1}{\sigma^2} \lambda_l)
\end{equation}
where $\lambda_l$ are the eigenvalues of $\frac{1}{m} {\bf H}{\bf
H}^{H}$. The problem consists therefore of estimating the
eigenvalues of $\frac{1}{m} {\bf H}{\bf H}^{H}$ based on few
observations $\hat{\bf H}_i$, which is paramount for modelling
purposes. Note that the capacity expression supposes that the
channel is perfectly known at the receiver and not at the
transmitter. In practice, with the noise impairment, the channel
will never be estimated perfectly and therefore expression
(\ref{channelcapacitydef}) is not achievable. However, for MIMO
modelling purposes, for which the capacity is often the matching
metric, one needs to compare the capacity of the model with
expression (\ref{channelcapacitydef}).

There are different methods actually used for channel capacity
estimation~\cite{channelest1,channelest2,channelest3,channelest4}. 
Usual methods discard, through an ad-hoc threshold procedure, all
channels $\hat{\bf H}_i$ for which the channel to noise ratio
($\frac{1}{\sigma^2 } tr_n({\bf H}{\bf H}^{H})$) is lower than a
threshold and then compute
\[
  \tilde{C}(\sigma^2) = \frac{1}{n} \log_2 \det \left( {\bf I}_n + \frac{1}{m\sigma^2} (\frac{1}{M} \sum_{i=1}^M \hat{{\bf H}}_i) (\frac{1}{M} \sum_{i=1}^M \hat{{\bf H}}_i)^H) \right)
\]
where $M \leq L$ is the number of channels having a signal to noise
ratio higher than the threshold. One of the drawbacks of this method
is that one will not analyze the true capacity but only the capacity
of the "good channels". Moreover, one has to limit the channel
measurement campaign (in order to have enough channels higher than
the threshold) only to regions which are close (in terms of actual
distance) enough to the base station.

Other methods, in order to have a capacity estimation at a given
signal to noise ratio (different from the measured one with noise
variance $\sigma^2$), normalize each channel realization $\hat{\bf
H}_i$ and then compute for a different value of the noise variance
$\sigma_1^2$ (for example $10dB$) the capacity estimate
$\tilde{C}(\sigma_1^2)$. In the case where $\sigma^2$ is high and
$\sigma_1^2$ is low, one usually finds a high capacity estimate
as one measures only the noise, which is known to have a high
multiplexing gain.

In this contribution, we will provide a neat framework, based on
free deconvolution, for channel capacity estimation that circumvents
all the previous drawbacks. Moreover, we will deal with model
(\ref{model2}), for which no solution has been provided in the
literature so far.

\section{Framework for free convolution} \label{framework}
Free probability~\cite{book:hiaipetz} theory has grown into an
entire field of research through the pioneering work of Voiculescu
in the 1980's. Free probability introduces an analogy to the
concept of independence from classical
probability, which can be used for non-commutative random variables
like matrices. These more general random variables are elements in
what is  called a  {\em noncommutative probability space}. This can
be defined by a pair $(A,\phi)$, where $A$ is a unital {\em
$\ast$-algebra} with unit $I$, and $\phi$ is a normalized (i.e.
$\phi(I)=1$) linear functional on $A$. The elements of $A$ are
called random variables. In all our examples, $A$ will consist of
$n\times n$ matrices or random matrices. For matrices, $\phi$ will be $tr_n$.
The unit in these $\ast$-algebras is the $n\times n$ identity matrix
${\bf I}_n$. The analogy to independence is called freeness:
\begin{definition} \label{freedef}
  A family of unital $\ast$-subalgebras
  $(A_i)_{i\in I}$ will be called a free family if
\begin{equation} \label{freeeq}
       \left\lbrace  \begin{matrix}
           a_j\in A_{i_j} \\
           i_1\neq i_2,i_2\neq i_3,\cdots ,i_{n-1}\neq i_n \\
           \phi(a_1)=\phi(a_2)=\cdots =\phi(a_n)=0 \end{matrix}
       \right\rbrace
  \Rightarrow \phi(a_1\cdots a_n)=0.
\end{equation}
  A family of random variables $a_i$ are said to be free if the algebras they generate form a free family.
\end{definition}

When restricting $A$ to spaces such as matrices, or functions with bounded support, it is clear that
the moments of $a$ uniquely identify a probability measure, here called $\nu_a$, such that $\phi(a^k) = \int x^k d\nu_a(x)$.
In such spaces, the distributions of $a_1+a_2$ and $a_1a_2$ give us two new probability measures, which depend only on the
probability measures associated with $a_1$, $a_2$ when these are free.
Therefore we can define two operations on the set of probability measures:
{\em Additive free convolution} $\eta_1\boxplus\eta_2$
for the sum of free random variables, and
{\em multiplicative free convolution} $\eta_1\boxtimes\eta_2$
for the product of free random variables. These operations can in many cases be
used to predict the spectrum of sums or products of large random matrices:
If $a_{1n}$ has an eigenvalue
distribution which approaches $\eta_1$ and $a_{2n}$ has an
eigenvalue distribution which approaches $\eta_2$, then in many cases the
eigenvalue distribution of $a_{1n}+a_{2n}$ approaches
$\eta_1\boxplus\eta_2$.

One important probability measure is the
Mar\u{c}henko Pastur law $\mu_c$~\cite{book:tulinoverdu},
which has the density
\begin{equation} \label{mpdensity}
  f^{\mu_c}(x) = (1-\frac{1}{c})^+ \delta_0(x) + \frac{\sqrt{(x-a)^+(b-x)^+}}{2\pi cx},
\end{equation}
where $(z)^+ =\mbox{max}(0,z)$, $a=(1-\sqrt{c})^2$, $b=(1+\sqrt{c})^2$, and $\delta_0(x)$ is dirac measure (point mass)
at $0$. According to the notation in~\cite{book:comblect}, 
$\mu_c$ is also the free Poisson distribution with rate $\frac{1}{c}$ and jump size $c$.
We will need the following formulas for the first moments of
the Mar\u{c}henko Pastur law:
\begin{equation} \label{firstthreemomentsmp}
\begin{array}{lll}
\int x f^{\mu_c}(x) dx   &=& 1\\
\int x^2 f^{\mu_c}(x) dx &=& c+1\\
\int x^3 f^{\mu_c}(x) dx &=& c^2 + 3c + 1\\
\int x^4 f^{\mu_c}(x) dx &=& c^3 + 6c^2 + 6c + 1.
\end{array}
\end{equation}
(\ref{firstthreemomentsmp}) follows immediately from applying what is called the {\em moment-cumulant formula}~\cite{book:comblect},
to the {\em free cumulants}~\cite{book:comblect} of the Mar\u{c}henko Pastur law $\mu_c$.
The (free) cumulants of the Mar\u{c}henko Pastur law are $1,c,c^2,c^3,...$~\cite{eurecom:multfreeconv}.
Cumulants and the moment-cumulant formula in free probability have analogous concepts in classical probability.

$\mu_c$ describes asymptotic eigenvalue distributions of {\em
Wishart matrices}, i.e. matrices on the form $\frac{1}{N} {\bf
R}{\bf R}^{H}$, with ${\bf R}$ an $n\times N$ random matrix with
independent standard complex Gaussian entries, and $\frac{n}{N}\rightarrow
c$. This can be seen from the following result, where the difference
from (\ref{firstthreemomentsmp}) vanishes when $N\rightarrow\infty$:
\begin{proposition} \label{propgaussian}
Let ${\bf X}_n$ be a complex standard Gaussian $n\times N$ matrix,
and set $c=\frac{n}{N}$. Then
\begin{equation} \label{firstthreemoments}
\begin{array}{lll}
E\left[ tr_n\left( \frac{1}{N} {\bf X}_n {\bf X}_n^H \right) \right]                 &=& 1\\
E\left[ tr_n\left( \left( \frac{1}{N} {\bf X}_n {\bf X}_n^H \right)^2 \right) \right]&=& c+1\\
E\left[ tr_n\left( \left( \frac{1}{N} {\bf X}_n {\bf X}_n^H \right)^3 \right) \right]&=& c^2 + 3c + 1 + \frac{1}{N^2}\\
E\left[ tr_n\left( \left( \frac{1}{N} {\bf X}_n {\bf X}_n^H
\right)^4 \right) \right]&=& c^3 + 6c^2 + 6c + 1 +
\frac{5(1+c)}{N^2}.
\end{array}
\end{equation}
\end{proposition}
This will be useful later on when we compute mixed moments of Gaussian and deterministic matrices. 
The proof of proposition~\ref{propgaussian} is given in appendix~\ref{appendixc0}.

We will also find it useful to introduce the concept of {\em multiplicative free deconvolution}:
Given probability measures $\eta$ and $\eta_2$.
When there is a unique probability measure $\eta_1$ such that
$\eta = \eta_1 \boxtimes \eta_2$,
we will write $\eta_1 = \eta \boxslash \eta_2$,
and say that $\eta_1$ is the multiplicative free deconvolution of $\eta$ with $\eta_2$.
There is no reason why a probability measure should have a unique deconvolution, and whether one exists at all depends highly on the
probability measure $\eta_2$ which we deconvolve with.
This will not be a problem for our purposes:
First of all, we will only have need for multiplicative free deconvolution with $\mu_c$,
and only in order to find the moments of the channel matrix. 
The problem of a unique deconvolution is therefore addressed by an existing algorithm for free deconvolution~\cite{eurecom:freedeconvinftheory}, 
which finds unique moments of $\eta\boxslash\mu_c$ (as long as the first moments of $\eta$ is nonzero).

We will need the following definitions:
\begin{definition}
By the {\em empirical eigenvalue distribution}
of an $n\times n$ random matrix ${\bf X}$ we mean the random atomic measure
\[
  \frac{1}{n} \left( \delta_{\lambda_1({\bf X})} + \cdots + \delta_{\lambda_n({\bf X})} \right),
\]
where $\lambda_1({\bf X}),...,\lambda_n({\bf X})$ are the (random) eigenvalues of ${\bf X}$.
\end{definition}
\begin{definition}
A sequence of random variables $a_{n1},a_{n2},...$ in probability spaces $(A_n,\phi_n)$ is said to converge in distribution if,
for any $m_1,...,m_r\in\N$, $k_1,...,k_r\in\{ 1,2,...\}$,
we have that the limit $\phi_n( a_{nk_1}^{m_1}\cdots a_{nk_r}^{m_r})$ exists as $n\rightarrow\infty$.
\end{definition}

To make the connection between models (\ref{model1}), (\ref{model2}) and model (\ref{system1}),
we need the following result~\cite{eurecom:multfreeconv}:
\begin{theorem} \label{teo1d}
  Assume that the empirical eigenvalue distribution of ${\bf \Gamma}_n=\frac{1}{N}{\bf R}_n{\bf R}_n^{H}$
  converges in distribution almost surely to a compactly supported probability measure $\eta_{\Gamma}$.
  Then we have that the empirical eigenvalue distribution of ${\bf W}_n$ also converges in distribution almost
  surely to a compactly supported probability measure $\eta_W$ uniquely identified by
  \begin{equation} \label{starteq}
    \eta_W \boxslash \mu_c = (\eta_{\Gamma} \boxslash \mu_c) \boxplus \delta_{\sigma^2},
  \end{equation}
  where $\delta_{\sigma^2}$ is dirac measure (point mass) at $\sigma^2$.
\end{theorem}

Theorem~\ref{teo1d} can also be re-stated (through deconvolution) as
\[
  \eta_{W} = \left( (\eta_{\Gamma} \boxslash \mu_c) \boxplus \delta_{\sigma^2} \right) \boxtimes \mu_c.
\]

When we have $L$ observations $\hat{\bf H}_i$ in a MIMO system as in (\ref{model1}) or (\ref{model2}),
we will form the $n\times mL$ random matrices
\begin{eqnarray}
  \hat{{\bf H}}_{1...L} = {\bf  H}_{1...L} + \frac{\sigma}{\sqrt{L}} {\bf X}_{1...L}
\end{eqnarray}
with
\[
  \hat{{\bf H}}_{1...L} =\frac{1}{\sqrt{L}} \left[\hat{{\bf H}}_1, \hat{{\bf H}}_2,...,\hat{{\bf H}}_L\right],
\]
\[
  {\bf H}_{1...L} = \frac{1}{\sqrt{L}} \left[{\bf D}_i^{r} {\bf  H} {\bf D}_i^{t},{\bf D}_i^{r} {\bf  H} {\bf D}_i^{t},...,{\bf D}_i^{r} {\bf  H} {\bf D}_i^{t}\right],
\]
\[
  {\bf X}_{1...L} =   \left[{\bf X}_{1},{\bf X}_{2},...,{\bf X}_L\right].
\]
This is the way we will stack the observations in this paper. It is only one of many possible stackings. 
A stacking where the ratio between the number of rows and the number of columns converges to a quantity between $0$ and $1$ 
would allow us to use theorem~\ref{teo1d} (which implicitly assumes $0<c<1$) directly to conclude almost sure convergence, which again would 
help us to conclude that the introduced capacity estimators are asymptotically unbiased. 
Such a stacking can also reduce the variance of the estimators. 
Even though the stacking considered here may not give the lowest variance, and may not give almost sure convergence, 
we show that its variance converges to $0$ and provides asymptotic unbiasedness for the corresponding capacity estimator. 

For the case $L=1$, the formula
\begin{equation} \label{usethis}
  tr_n\left( \left( {\bf D}_1^{r} {\bf H} {\bf D}_1^{t} \left( {\bf D}_1^{r} {\bf H} {\bf D}_1^{t} \right)^H \right)^j \right)
  =
  tr_n\left( \left( {\bf H}{\bf H}^H \right)^j \right)
\end{equation}
can be combined with theorem~\ref{teo1d} to give the approximation
\begin{equation} \label{decthis}
  \nu_{ \frac{1}{m}\hat{\bf H}_1\hat{\bf H}_1^{H} } \boxslash \mu_{\frac{n}{m}}
  \approx
  \left( \nu_{\frac{1}{m} {\bf H}{\bf H}^{H}} \boxslash \mu_{\frac{n}{m}} \right) \boxplus \delta_{\sigma^2}.
\end{equation}
for a single observation. This approximation works well when $n$ is large.
For many observations, note that ${\bf H}_{1...L} {\bf H}_{1...L}^{H} = {\bf H}{\bf H}^{H}$
when there is no phase off-set and phase drift, so that the approximation
\begin{equation} \label{decthis2}
  \nu_{ \frac{1}{m} \hat{\bf H}_{1...L}\hat{\bf H}_{1...L}^{H} } \boxslash \mu_{\frac{n}{mL}}
  \approx
  \left( \nu_{\frac{1}{m} {\bf H}{\bf H}^{H}} \boxslash \mu_{\frac{n}{mL}} \right) \boxplus \delta_{\sigma^2}
\end{equation}
applies and generalizes (\ref{decthis}).
The ratio between the number of rows and columns in the matrices ${\bf H}_{1...L}, {\bf X}_{1...L}$ and $\hat{\bf H}_{1...L}$ is $c=\frac{n}{mL}$,
considering the horizontal stacking of the observations in a larger matrix.
It is only this stacking which will be considered in this paper.

When phase off-set and phase drift are added, it is much harder to
adapt theorem~\ref{teo1d} to produce the moments of $\frac{1}{m}
{\bf H}{\bf H}^{H}$. The reason is that theorem~\ref{teo1d} really
helps us to find the moments of $\frac{1}{m} {\bf  H}_{1...L} {\bf
H}_{1...L}^H$. In the case without phase off-set and phase drift,
this is enough since these moments are equal to the moments of
$\frac{1}{m}{\bf H}{\bf H}^H$. However, equality between these
moments does not hold when phase off-set and phase drift are added.
A procedure for converting between these moments may exist, but
seems to be rather complex, and will not be dealt with here. In
section~\ref{newestimators}, we will instead define an estimator for
the channel capacity which does not stack observations into the
matrix ${\bf  H}_{1...L}$ at all. Instead, an estimation will be
performed for each observation, taking the mean of all the estimates
at the end.

\section{New estimators for channel capacity} \label{newestimators}
In this section, two new channel capacity estimators are defined.
First, a {\em free probability based estimator} is introduced, which
(for model (\ref{model1})) will be shown to be asymptotically unbiased w.r.t. the number of
observations. Then, by slightly modifying the free probability based
estimator, we will construct what we call the {\em Gaussian matrix mean based capacity
estimator}. This estimator will be shown, for model (\ref{model1}) and (\ref{model2}), to give unbiased estimates of the moments of the channel matrix for any
number of observations. The computational complexity
for the two estimators lies in the computation of eigenvalues and moments of the matrix ${\bf H}{\bf H}^{H}$,
in addition to computing the free (de)convolution in terms of moments. For the matrix ranks considered here,
free (de)convolution requires few computations.
The complexity in the computation of eigenvalues and moments of the matrix ${\bf H}{\bf H}^{H}$ grows with $n$ (the number of receiving antennas),
which is small in this paper.
The computational complexity in the estimators grows slowly with the number of observations,
since the dimensions of $\hat{\bf H}_{1...L}\hat{\bf H}_{1...L}^{H}$ does not grow with $L$.

The two estimators are stated as estimators for the lower order moments of $\frac{1}{m} {\bf H}{\bf H}^{H}$.
Under the assumption that this matrix has limited rank (such as $\leq 4$ here),
estimators for lower order moments can be used to define estimators for the channel capacity,
since the capacity can be written as a function of the $r$ lowest moments when the matrix has rank $r$, as explained below.

\subsection{The free probability based capacity estimator}
The free probability based estimator is defined as follows:
\begin{definition} \label{deffreeestimator}
The free probability based estimator for the capacity of a channel with channel matrix ${\bf H}$ of rank $r$, denoted $C_f$,
is computed through the following steps:
\begin{enumerate}
  \item Compute the first $r$ moments $\hat{h}_1,...,\hat{h}_r$
    of the sample covariance matrix $\frac{1}{m} \hat{\bf H}_{1...L}\hat{\bf H}_{1...L}^{H}$
    (i.e. compute $\hat{h}_j = tr_n \left( \left( \frac{1}{m} \hat{\bf H}_{1...L}\hat{\bf H}_{1...L}^{H}  \right)^j \right)$ for $1\leq j\leq r$),
  \item use (\ref{decthis2}) to estimate the first $r$ moments $h_{f1},...,h_{fr}$ of $\frac{1}{m} {\bf H}{\bf H}^{H}$,
  \item estimate the $r$ nonzero eigenvalues $\lambda_1,...,\lambda_r$ of $\frac{1}{m} {\bf H}{\bf H}^{H}$ from $h_{f1},...,h_{fr}$.
    Substitute these in (\ref{channelcapacitydef}).
\end{enumerate}
We also call $h_{f1},...,h_{fr}$ the free probability based estimators for the $r$ first moments of  $\frac{1}{m} {\bf H}{\bf H}^{H}$.
\end{definition}

Steps 2 and 3 in definition~\ref{deffreeestimator} need some elaboration. To
address step 3, consider the case of a rank $3$ channel matrix. For
such channel matrices, only the lowest three moments $h_1$, $h_2$,
$h_3$ of $\frac{1}{m} {\bf H}{\bf H}^{H}$ need to be estimated in
order to estimate the eigenvalues. To
see this, first write
\begin{equation} \label{capacityeq}
\begin{array}{lll}
  C &=& \frac{1}{n}\log_2 \det\left( {\bf I}_n+\frac{1}{m \sigma^2} {\bf H}{\bf H}^{H} \right)\\
    &=& \frac{1}{n}\log_2 \left( \left( 1+\frac{1}{\sigma^2}\lambda_1 \right) \left( 1+\frac{1}{\sigma^2}\lambda_2 \right) \left( 1+\frac{1}{\sigma^2}\lambda_3 \right) \right),
\end{array}
\end{equation}
where $\lambda_1$, $\lambda_2$ and $\lambda_3$ are the three
non-zero eigenvalues of $\frac{1}{m}{\bf H}{\bf H}^{H}$. This
quantity can easily be calculated from the elementary symmetric
polynomials
\[
\begin{array}{lll}
  \Pi_1(\lambda_1, \lambda_2, \lambda_3) &=& \lambda_1 + \lambda_2 + \lambda_3 \\
  \Pi_2(\lambda_1, \lambda_2, \lambda_3) &=& \lambda_1 \lambda_2 + \lambda_2 \lambda_3+ \lambda_1 \lambda_3 \\
  \Pi_3(\lambda_1, \lambda_2, \lambda_3) &=& \lambda_1 \lambda_2 \lambda_3.
\end{array}
\]
by observing that
\[
  \left( 1+\frac{1}{\sigma^2}\lambda_1 \right) \left( 1+\frac{1}{\sigma^2}\lambda_2 \right) \left( 1+\frac{1}{\sigma^2}\lambda_3 \right)
\]
can be written as
\begin{equation} \label{girardeq1}
  1
  +
  \frac{1}{\sigma^2} \Pi_1(\lambda_1, \lambda_2, \lambda_3)
  +
  \frac{1}{\sigma^4} \Pi_2(\lambda_1, \lambda_2, \lambda_3)
  +
  \frac{1}{\sigma^6} \Pi_3(\lambda_1, \lambda_2, \lambda_3).
\end{equation}
$\Pi_1(\lambda_1, \lambda_2, \lambda_3)$ can in turn be calculated from the power polynomials
\[
\begin{array}{lll}
  S_1(\lambda_1 ,\lambda_2 ,\lambda_3) &=& \lambda_1 + \lambda_2 + \lambda_3       = n tr_n\left( \frac{1}{m}{\bf H}{\bf H}^{H} \right) \\
  S_2(\lambda_1 ,\lambda_2 ,\lambda_3) &=& \lambda_1^2 + \lambda_2^2 + \lambda_3^2 = n tr_n\left( \left( \frac{1}{m}{\bf H}{\bf H}^{H} \right)^2 \right)\\
  S_3(\lambda_1 ,\lambda_2 ,\lambda_3) &=& \lambda_1^3 + \lambda_2^3 + \lambda_3^3 = n tr_n\left( \left( \frac{1}{m}{\bf H}{\bf H}^{H} \right)^3 \right)
\end{array}
\]
by using the {\em Newton-Girard
formulas}~\cite{book:programmingmath}, which for the three first
moments take the form $\Pi_1 = S_1$, $\Pi_2 = \frac{1}{2} \left( S_1^2 - S_2 \right)$ and $\Pi_3 = \frac{1}{6} \left( S_1^3 - 7S_1S_2 +2S_3 \right)$. 
If the channel matrix has a higher rank $r$, similar reasoning can be used to conclude that the first $r$ moments need to be estimated.
In the simulations, the eigenvalues themselves are never computed,
since computation of the moments and the Newton-Girard formulas make this unnecessary.

To address step 2 in definition~\ref{deffreeestimator},
a Matlab implementation~\cite{eurecom:capacityimpl} which performs free (de)convolution in terms of moments as described in~\cite{eurecom:freedeconvinftheory}
was developed and used for the simulations in this paper.
Free (de)convolution is computationally expensive for higher order moments only:
For the first four moments, step 2 in definition~\ref{deffreeestimator} is equivalent to the following:

\begin{proposition} \label{propfreeestimator}
Let $\hat{h}_1,\hat{h}_2,\hat{h}_3,\hat{h}_4$ and $h_{f1},h_{f2},h_{f3},h_{f4}$ be as in definition~\ref{deffreeestimator}. Then
\begin{equation} \label{freeestimator}
\begin{array}{lll}
  \hat{h}_1 &=& h_{f1} + \sigma^2\\
  \hat{h}_2 &=& h_{f2} + 2{\sigma}^2 (1+c) h_{f1} + {\sigma}^4 (1 + c) \\
  \hat{h}_3 &=& h_{f3} + 3{\sigma}^2(1+c) h_{f2} + 3{\sigma}^2 c h_{f1}^2 \\
            & & + 3{\sigma}^4 \left( c^2+3c+1 \right) h_{f1} \\
            & & + {\sigma}^6 \left( c^2+3c+1 \right) \\
  \hat{h}_4 &=& h_{f4} + 4{\sigma}^2(1+c) h_{f3} + 8{\sigma}^2 c h_{f2} h_{f1} \\
            & & + {\sigma}^4 (6c^2+16c+6) h_{f2} \\
            & & + 14{\sigma}^4 c(1+c) h_{f1}^2 \\
            & & + 4{\sigma}^6 (c^3+ 6c^2+6c+1) h_{f1} \\
            & & + {\sigma}^8 \left( c^3 + 6c^2 + 6c + 1 \right),
\end{array}
\end{equation}
where $c=\frac{n}{mL}$.
\end{proposition}

The proof of proposition~\ref{propfreeestimator} can be found in
appendix~\ref{appendixb}. The following is the main result
on the free probability based estimator, and covers the
different cases for bias and asymptotic bias w.r.t. number of
observations or antennas.
\begin{theorem} \label{teofreeestimatormain}
  For $L=1$ observation, the following holds for both models (\ref{model2}) and (\ref{model1}):
  \begin{enumerate}
    \item $h_{f1}$ and $h_{f2}$ are unbiased. $h_{f3}$ and $h_{f4}$ are biased, with the bias of $h_{f3}$ given by
      \[
        -\frac{3{\sigma}^4 tr_n\left( \frac{1}{m}{\bf H}{\bf H}^H \right) + {\sigma}^6}{m^2}.
      \]
      In particular $h_{f3}$ and $h_{f4}$ are asymptotically unbiased when $m\rightarrow\infty$ (with $n,L$ kept fixed), i.e.
      \[
        \lim_{m\rightarrow\infty} E(h_{fj}) = tr_n \left( \left( \frac{1}{m} {\bf H}{\bf H}^{H} \right)^j \right) \mbox{, } 3 \leq j \leq 4.
      \]
    \item $C_f$ is asymptotically unbiased when $m\rightarrow\infty$ (with $n,L$ kept fixed) 
      and $\frac{1}{m} {\bf H}{\bf H}^{H}$ has rank $\leq 4$, i.e. $\lim_{m\rightarrow\infty} C_f = C$.
  \end{enumerate}
  For any number of observations $L$ with model (\ref{model1}), the following holds:
  \begin{enumerate}
    \item $h_{f1}$ and $h_{f2}$ are unbiased. $h_{f3}$ and $h_{f4}$ are biased, with the bias of $h_{f3}$ given by
      \[
        -\frac{3{\sigma}^4 tr_n\left( \frac{1}{m}{\bf H}{\bf H}^H \right) + {\sigma}^6}{m^2L^2}.
      \]
      In particular $h_{f3}$ and $h_{f4}$ are asymptotically unbiased when either $m\rightarrow\infty$ or $L\rightarrow\infty$ (with the other kept fixed), i.e.
      \[
        \lim_{m\rightarrow\infty} E(h_{fj})
        =
        \lim_{L\rightarrow\infty} E(h_{fj})
        =
        tr_n \left( \left( \frac{1}{m} {\bf H}{\bf H}^{H} \right)^j \right)
      \]
      for $3 \leq j \leq 4$.
    \item $C_f$ is asymptotically unbiased when either 
      $m\rightarrow\infty$ (with $n,L$ kept fixed),
      or $L\rightarrow\infty$ (with $m,n$ kept fixed)
      and $\frac{1}{m} {\bf H}{\bf H}^{H}$ has rank $\leq 4$, i.e. $\lim_{m\rightarrow\infty} C_f = \lim_{L\rightarrow\infty} C_f = C$.
  \end{enumerate}
\end{theorem}

The proof of theorem~\ref{teofreeestimatormain} can be found in appendix~\ref{appendixc}.
The bias in theorem~\ref{teofreeestimatormain} motivates the definition of the estimator of the next section.
The free probability based estimator performs estimation as if the Gaussian random matrices and deterministic matrices involved were free.
It turns out that these matrices are only {\em asymptotically free}~\cite{book:hiaipetz},
which explains why there is a bias involved, and why the bias decreases as the matrix dimensions increase.

\subsection{The Gaussian matrix mean based capacity estimator}
The expression for the Gaussian matrix mean based capacity estimator is motivated from
computing expected values of mixed moments of Gaussian and deterministic matrices (lemma~\ref{lemgeneralized}). 
This results in expressions slightly different from (\ref{freeestimator}).
We will show that the Gaussian matrix mean based estimator can be used for channel capacity estimation in certain systems
where the free probability based estimator fails.
The definition of the Gaussian matrix mean based capacity estimator is as follows for matrices of rank $\leq 4$:

\begin{definition} \label{defunbiasedestimator}
The Gaussian matrix mean based estimator for the capacity of a channel with channel matrix ${\bf H}$ of rank $r\leq 4$, denoted $C_G$,
is defined through the following steps:
\begin{enumerate}
  \item For each observation, perform the following
    \begin{enumerate}
      \item Compute the first $r$ moments $\hat{h}_{i1},...,\hat{h}_{ir}$ of the sample covariance matrix $\frac{1}{m} \hat{\bf H}_i\hat{\bf H}_i^{H}$
        (i.e. compute $\hat{h}_{ij} = tr_n \left( \left( \frac{1}{m} \hat{\bf H}_i\hat{\bf H}_i^{H}  \right)^j \right)$ for $1\leq j\leq r$),
      \item find estimates $h_{i1},h_{i2},h_{i3},h_{i4}$ of the first four moments of $\frac{1}{m} {\bf H} {\bf H}^{H}$ by solving
        \begin{equation} \label{unbiasedestimator}
          \begin{array}{lll}
            \hat{h}_{i1} &=& h_{i1} + \sigma^2 \\
            \hat{h}_{i2} &=& h_{i2} + 2{\sigma}^2 (1+c) h_{i1} + {\sigma}^4 (1 + c) \\
            \hat{h}_{i3} &=& h_{i3} + 3{\sigma}^2(1+c) h_{i2} + 3{\sigma}^2 c h_{i1}^2 \\
                         & & + 3{\sigma}^4 \left( c^2+3c+1+\frac{1}{m^2}\right) h_{i1} \\
                         & & + {\sigma}^6 \left( c^2+3c+1+\frac{1}{m^2}\right)\\
            \hat{h}_{i4} &=& h_{i4} + 4{\sigma}^2(1+c) h_{i3} + 8{\sigma}^2 c h_{i2} h_{i1} \\ 
                         & & + {\sigma}^4 (6c^2+16c+6+\frac{16}{m^2}) h_{i2} \\
                         & & + 14{\sigma}^4 c(1+c) h_{i1}^2 \\
                         & & + 4{\sigma}^6 (c^3+ 6c^2+6c+1+\frac{5(c+1)}{m^2}) h_{i1} \\
                         & & + {\sigma}^8 \left( c^3 + 6c^2 + 6c + 1 + \frac{5(c+1)}{m^2}\right),
          \end{array}
        \end{equation}
        where $c=\frac{n}{m}$,
    \end{enumerate}
    Form the estimates $h_{uj} = \frac{1}{L} \sum_{i=1}^L h_{ij}$, $1\leq j\leq r$, of the first moments of $\frac{1}{m} {\bf H} {\bf H}^{H}$,
  \item estimate the $r$ nonzero eigenvalues $\lambda_1,...\lambda_r$ of $\frac{1}{m} {\bf H}{\bf H}^{H}$ from $h_{u1},...,h_{ur}$.
    Substitute these in (\ref{channelcapacitydef}).
\end{enumerate}
We also call $h_{u1},...,h_{ur}$ the Gaussian matrix mean based estimators for the $r$ first moments of  $\frac{1}{m} {\bf H}{\bf H}^{H}$.
\end{definition}

While a Matlab implementation~\cite{eurecom:capacityimpl} of free (de)convolution is used for the free (de)convolution
in the free probability based estimator, the algorithm for the Gaussian matrix mean based capacity estimator used by the simulations in this paper
follows the steps in definition~\ref{defunbiasedestimator} directly.

Note that (\ref{unbiasedestimator}) resemble the formulas in (\ref{freeestimator}) when $c=\frac{n}{m}$.
$c=\frac{n}{m}$ is used in definition~\ref{defunbiasedestimator} since the observation matrices $\hat{\bf H}_i$
are not stacked together in a larger matrix in this case.
Instead, a mean is taken of all estimated moments in step 1 of the definition.
This is not an optimal procedure, and we use it only because it is hard to compute mixed
moments of matrices where observations $\hat{{\bf H}}_i$ of type (\ref{model2}) are
stacked together.

The following theorem is the main result on the Gaussian matrix mean based estimator,
and shows that it qualifies for it's name.

\begin{theorem} \label{teounbiasedestimatormain}
  For either model (\ref{model1}) or (\ref{model2}), the following holds:
  \begin{enumerate}
    \item The estimators $h_{u1},h_{u2},h_{u3},h_{u4}$ are unbiased, i.e.
      \[
        E(h_{uj}) = tr_n \left( \left( \frac{1}{m} {\bf H}{\bf H}^{H} \right)^j \right) \mbox{, } 1 \leq j \leq 4.
      \]
    \item $C_G$ is asymptotically unbiased as $m\rightarrow\infty$ (with $n,L$ kept fixed) 
      when $\frac{1}{m} {\bf H}{\bf H}^{H}$ has rank $\leq 4$, i.e. $\lim_{m\rightarrow\infty} C_G = C$.
    \item In the case of $L=1$ observation, $h_{f1} = h_{u1}$ and $h_{f2} = h_{u2}$.
      In particular, $C_f=C_G$ when $\frac{1}{m} {\bf H}{\bf H}^{H}$ has rank $\leq 2$.
  \end{enumerate}
\end{theorem}
The proof of theorem~\ref{teounbiasedestimatormain} can be found in appendix~\ref{appendixc}.

\subsection{Limitations of the two estimators}
We have chosen to define two estimators, since they have different limitations.

The most severe limitation of the Gaussian matrix mean based capacity estimator, the way it is defined, lies in the restriction on the rank.
This restriction is done to limit the complexity in the expression for the estimator. 
However, the computations in appendix~\ref{appendixc} should convince the reader that
capacity estimators with similar properties can be written down (however complex) for higher rank channels also.
Also, while the free probability based estimator has an algorithm~\cite{eurecom:freedeconvinftheory} for channel matrices of any rank, 
there is no reason why a similar algorithm can not be found for the Gaussian matrix mean based estimator also. 
The computations in appendix~\ref{appendixc} indicate that 
such an algorithm should be based solely on iteration through a finite set of partitions. 
How this can be done algorithmically is beyond the scope of this paper. 

For the free probability based estimator the limitation lies in the presence of phase off-set and phase drift (model (\ref{model2})):
When model (\ref{model2}) is used, the comments at the end of section~\ref{framework} make it clear that we lack a relation for obtaining the 
moments of  $\frac{1}{m} {\bf  H}_{1...L} {\bf H}_{1...L}^H$ from the moments of $\frac{1}{m}{\bf H}{\bf H}^H$.
Without such a relation, we also have no candidate for a capacity estimator 
(capacity estimators in this paper are motivated by first finding moment estimators). 
In conclusion, the stacking of observations performed by the free probability based estimator does not work for model (\ref{model2}). 
Only the Gaussian matrix mean based estimator can perform reliable capacity estimation for many observations with model (\ref{model2}). 
The second limitation of the free probability based estimator comes from the inherent bias in its deconvolution formulas (\ref{freeestimator}). 
The bias is only large when both $m$ and $L$ are small (see theorem~\ref{teofreeestimatormain}), so this point is less severe 
(however, channel matrices down to size $4\times 4$ occur in practice). 
The bias in the lower order moments is easily seen to affect capacity estimation from the following expansion of the capacity
\begin{equation} \label{series2}
\begin{array}{lll}
  C &=& \frac{1}{\ln 2} \sum_{k=1}^{\infty} \frac{(-1)^{k+1}m_k \rho^k}{k},
\end{array}
\end{equation}
which can be obtained from substituting the Taylor expansion
\begin{equation} \label{taylorseries}
  \log_2(1+t) = \frac{1}{\ln 2} \sum_{k=1}^{\infty} (-1)^{k+1}\frac{t^k}{k}
\end{equation}
into the definition of the capacity. 
Here $\rho = 1/\sigma^2$ is SNR, and $m_k$ are the moments of $\frac{1}{m}{\bf H}{\bf H}^H$. 
It is clear from (\ref{series2}) that, at least if we restrict to small $\rho$, 
the expression is dominated by the contribution from the first order moments. 
If $m$ is small we therefore first have a high relative error in the first moments after the deconvolution step, 
which will propagate to a high relative error in the capacity estimate for small $\rho$ due to (\ref{series2}). 
Thus, free probability based capacity estimation will work poorly for small $m,L$ and $\rho$. 
The same limitation is not present in the Gaussian matrix mean based estimator, since its moment estimators are unbiased. 

The limitation on the rank can in some cases be avoided, if we instead have some bounds on the eigenvalues:
If we instead knew that at most four of the eigenvalues are not "negligible", we could still use proposition~\ref{propfreeestimator} to
estimate the capacity. This follows from results on the continuity of multiplicative free convolution, which has been covered in~\cite{paper:vounbounded}.
Such continuity issues are also beyond the scope of this paper.

\section{Channel capacity estimation} \label{simulations}
Several candidates for channel capacity estimators for (\ref{model1}) have been used in the literature.
We will consider the following:
\begin{equation}
\begin{array}{l}
  C_1 = \frac{1}{nL} \sum_{i=1}^L \log_2 \det\left({\bf I}_n+\frac{1}{m \sigma^2}  \hat{\bf H}_i\hat{\bf H}_i^{H} \right)\\
  C_2 = \frac{1}{n} \log_2 \det \left( {\bf I}_n + \frac{1}{L\sigma^2m } \sum_{i=1}^L \hat{{\bf H}}_i \hat{{\bf H}}_i^H \right)\\
  C_3 = \frac{1}{n} \log_2 \det \left( {\bf I}_n + \frac{1}{\sigma^2 m } (\frac{1}{L} \sum_{i=1}^L \hat{{\bf H}}_i) (\frac{1}{L} \sum_{i=1}^L \hat{{\bf H}}_i)^H) \right)
\end{array}
\end{equation}
These will be compared with the free probability based ($C_f$) and the Gaussian matrix mean based ($C_G$) estimators.

\subsection{Channels without phase off-set and phase drift}
\begin{figure}
  \begin{center}
    \epsfig{figure=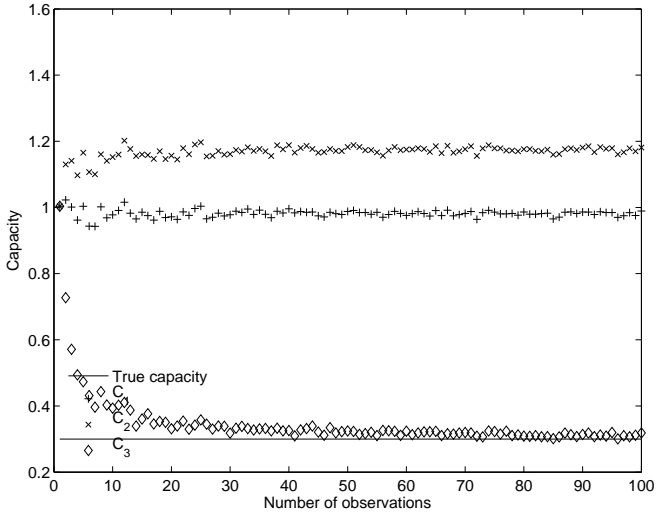,width=0.99\columnwidth}
  \end{center}
  \caption{Comparison of various classical capacity estimators for various number of observations, model (\ref{model1}).
           $\sigma^2 = 0.1$, $n=10$ receive antennas, $m=10$ transmit antennas. The rank of ${\bf H}$ was $3$.}\label{fig:sim1}
\end{figure}

In figure~\ref{fig:sim1}, $C_1$, $C_2$ and $C_3$ are compared for various number of
observations, with $\sigma^2 = 0.1$, and a $10\times 10$ channel
matrix of rank 3. It is seen that only the $C_3$ estimator gives
values close to the true capacity. The channel considered has no
phase drift or phase off-set. $C_1$ and $C_2$ are seen to have a
high bias.

\begin{figure}
  \begin{center}
    \epsfig{figure=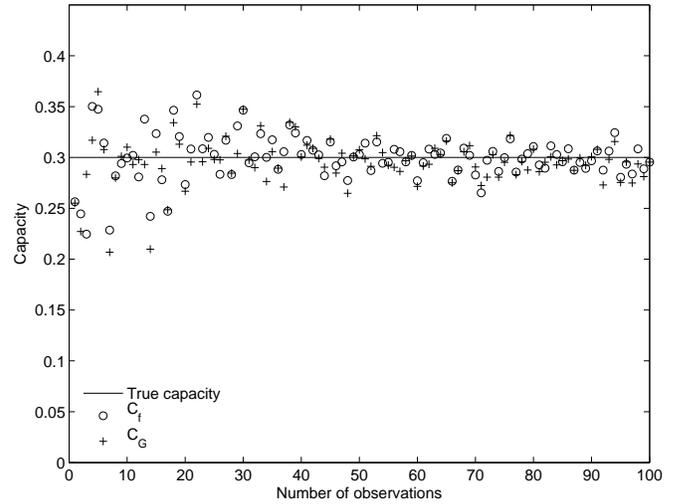,width=0.99\columnwidth}
  \end{center}
  \caption{Comparison of $C_f$ and $C_G$ for various number of observations, model (\ref{model1}).
           $\sigma^2 = 0.1$, $n=10$ receive antennas, $m=10$ transmit antennas. The rank of ${\bf H}$ was $3$.}\label{fig:sim3}
\end{figure}

In figure~\ref{fig:sim3}, the same $\sigma$ and channel matrix are
put to the test with the free-probability based and Gaussian matrix mean based
estimators for various number of observations. These give values
close to the true capacity. Both work better than $C_3$ for small
number of observations.

\begin{figure}
  \begin{center}
    \epsfig{figure=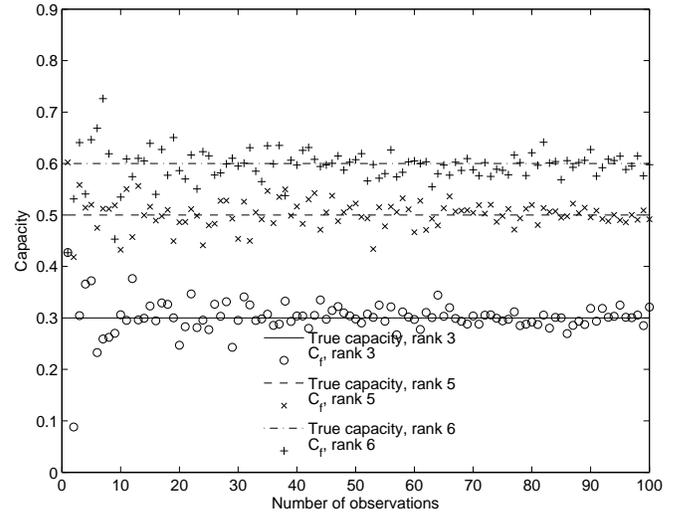,width=0.99\columnwidth}
  \end{center}
  \caption{$C_f$ for various number of observations, model (\ref{model1}).
           $\sigma^2 = 0.1$, $n=10$ receive antennas, $m=10$ transmit antennas. The rank of ${\bf H}$ was $3$, $5$ and $6$.}\label{fig:simrank}
\end{figure}

The free-probability based estimator converges faster (in terms of the number of observations) for lower rank channel matrices.
In figure~\ref{fig:simrank} we illustrate this for $10\times 10$ channel matrices of rank 3, 5 and 6.
Simulations show that for channel matrices of lower dimension (for instance $6\times 6$), we have slower convergence to the true capacity.

\subsection{Channels with phase off-set and phase drift}
\begin{figure}
  \begin{center}
    \epsfig{figure=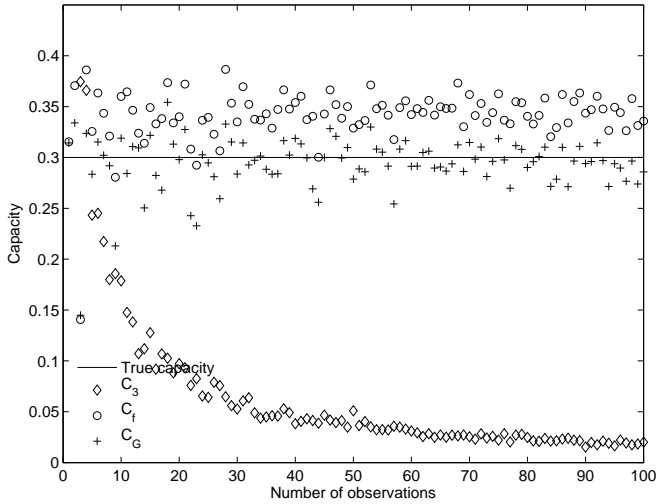,width=0.99\columnwidth}
  \end{center}
  \caption{Comparison of capacity estimators which worked for model (\ref{model1}) for increasing number of observations.
           Model (\ref{model2}) is used. $\sigma^2 = 0.1$, $n=10$ receive antennas, $m=10$ transmit antennas.
           The rank of ${\bf H}$ was $3$.}\label{fig:sim2}
\end{figure}

In figure~\ref{fig:sim2}, the $C_3$ estimator is compared with the free-probability based estimator, the Gaussian matrix mean based estimator and the true capacity,
for various number of observations, and with the same $\sigma$ and channel matrix as in figure~\ref{fig:sim1} and~\ref{fig:sim3}.
Phase off-set and phase drift have also been introduced.
In this case, the free-probability based estimator and the $C_3$-estimator seem to be biased.

\begin{figure}
  \begin{center}
    \epsfig{figure=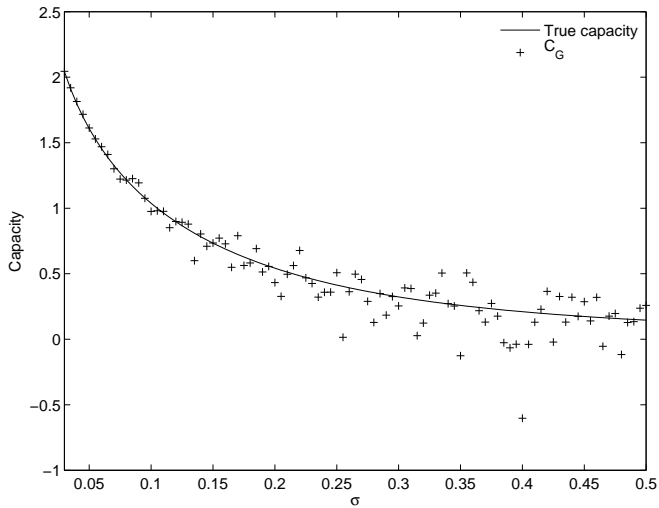,width=0.99\columnwidth}
  \end{center}
  \caption{$C_G$ for $L=1$ observation, $n=10$ receive antennas, $m=10$ transmit antennas, with varying values of $\sigma$.
           Model (\ref{model2}). The rank of ${\bf H}$ was $3$.} \label{fig:L1n10r3}
\end{figure}

In figure~\ref{fig:L1n10r3}, simulations have been performed for
various $\sigma$. Only $L=1$ observation was used, $n=10$ receive antennas, and $m=10$ transmit antennas. The
channel matrix has rank $3$. It is seen that the Gaussian matrix mean based capacity
estimator is very close to the true capacity, There are only small
deviations even if  one observation is present, which provides a
very good candidate for channel estimation in highly time-varying
environments. The deviations are higher for higher $\sigma$.

\begin{figure}
  \begin{center}
    \epsfig{figure=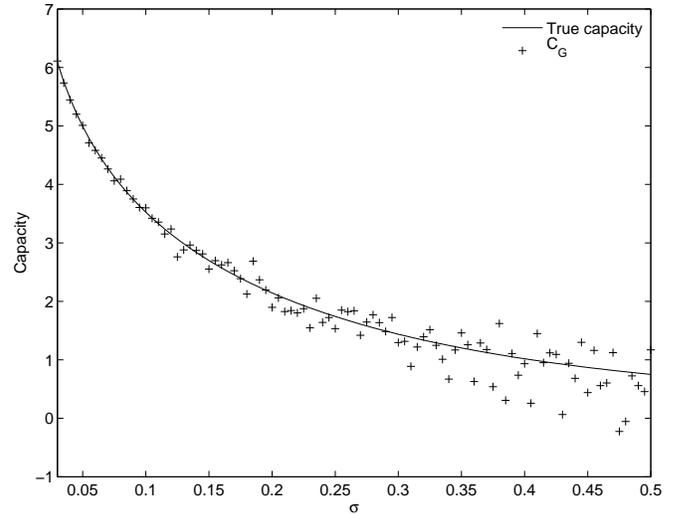,width=0.99\columnwidth}
  \end{center}
  \caption{$C_G$ for $L=1$ observation, $n=4$ receive antennas, $m=4$ transmit antennas, with varying values of $\sigma$.
           Model (\ref{model2}). The rank of ${\bf H}$ was $3$.} \label{fig:L1n4r3}
\end{figure}

\begin{figure}
  \begin{center}
    \epsfig{figure=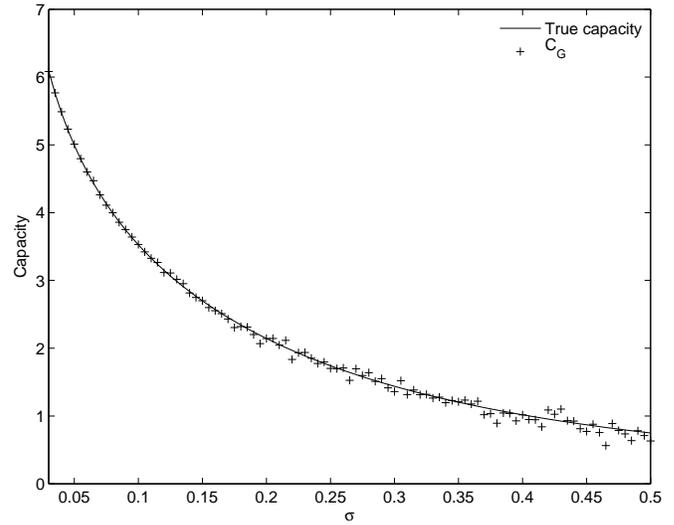,width=0.99\columnwidth}
  \end{center}
  \caption{$C_G$ for $L=10$ observations, $n=4$ receive antennas, $m=4$ transmit antennas, with varying values of $\sigma$.
           Model (\ref{model2}). The rank of ${\bf H}$ was $3$.} \label{fig:L10n4r3}
\end{figure}

In figure~\ref{fig:L1n4r3} we have also varied $\sigma$ and used
only one observation, but we have formed another rank $3$ matrix
with, $n=4$ receive antennas, $m=4$ transmit antennas. It is seen that the deviation from the true capacity is
much higher in this case.
We have in figure~\ref{fig:L10n4r3} increased the number of observations to $10$,
and used the same channel matrix.
It is seen that this decreases the deviation from the true capacity.

\begin{figure}
  \begin{center}
    \epsfig{figure=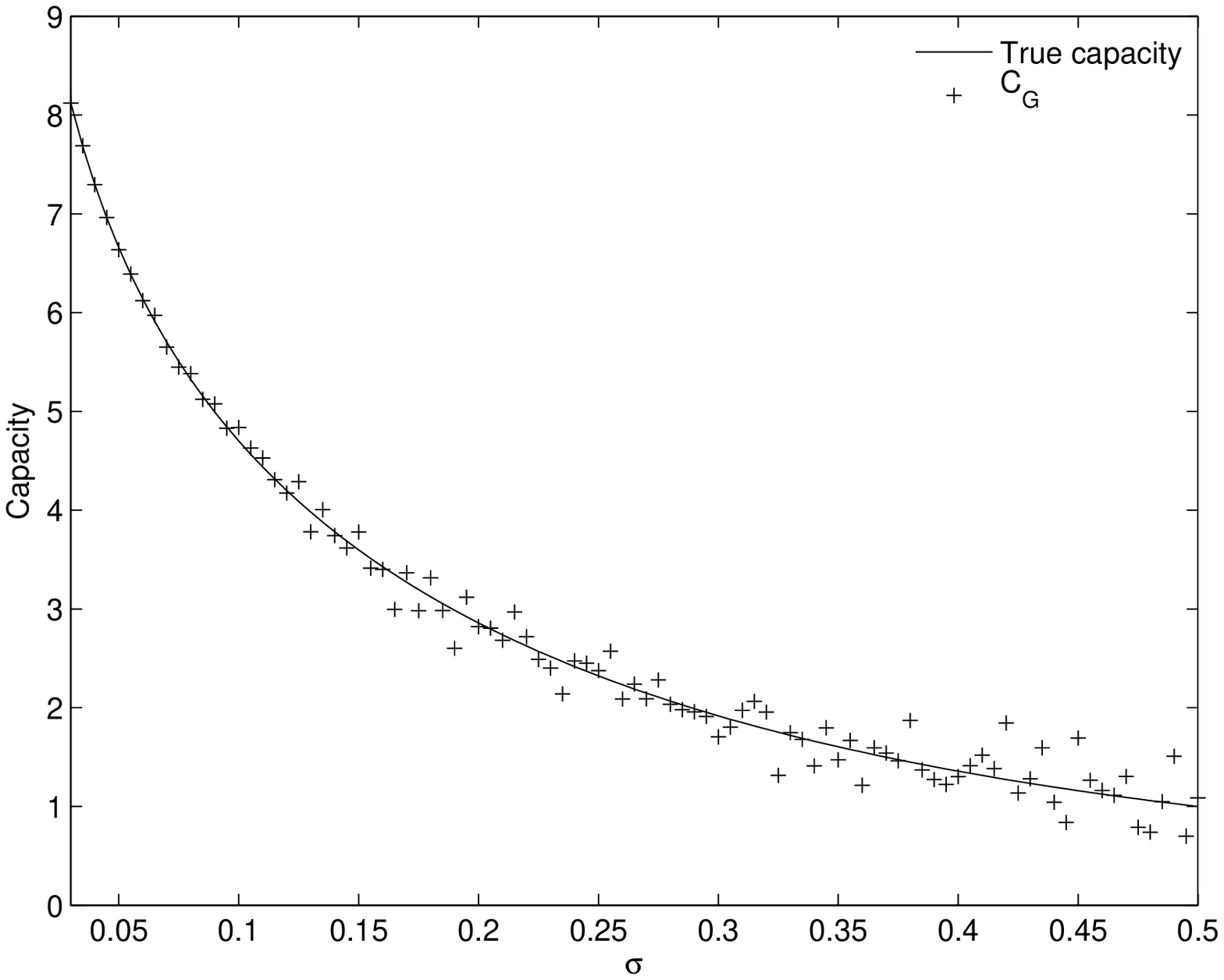,width=0.99\columnwidth}
  \end{center}
  \caption{$C_G$ for $L=1$ observation, $n=4$ receive antennas, $m=4$ transmit antennas, with varying values of $\sigma$.
           Model (\ref{model2}). The rank of ${\bf H}$ was $4$.} \label{fig:L1n4r4}
\end{figure}

\begin{figure}
  \begin{center}
    \epsfig{figure=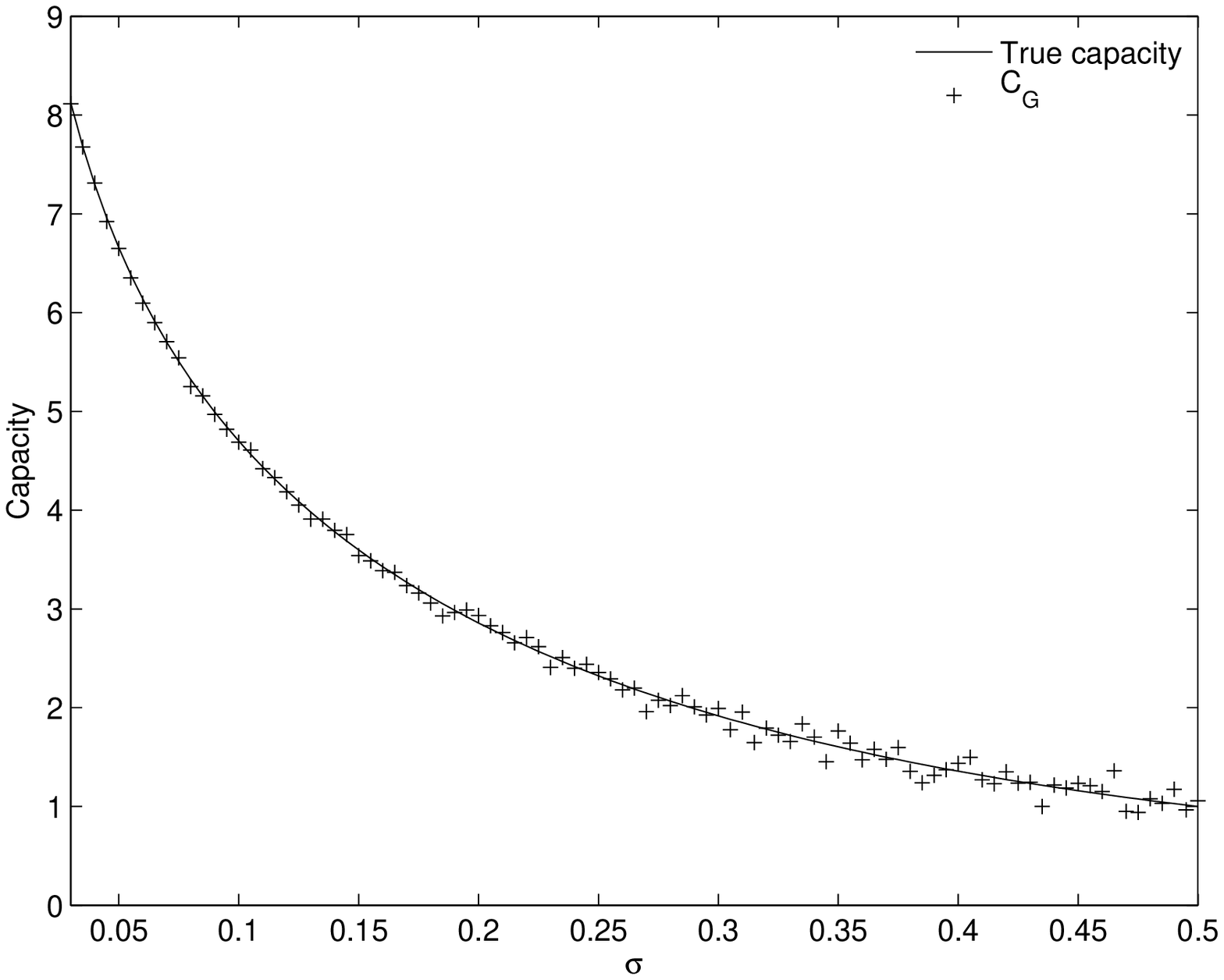,width=0.99\columnwidth}
  \end{center}
  \caption{$C_G$ for $L=10$ observations, $n=4$ receive antennas, $m=4$ transmit antennas, with varying values of $\sigma$.
           Model (\ref{model2}). The rank of ${\bf H}$ was $4$.} \label{fig:L10n4r4}
\end{figure}

Finally, let us use a channel matrix of rank $4$. In this case we
have to increase the number of observations even further to
accurately predict the channel capacity. In
figure~\ref{fig:L1n4r4}, Gaussian matrix mean based capacity estimation is performed
for a rank $4$ channel matrix with $n=4$ receive antennas, $m=4$ transmit antennas.
$1$ observation is performed. If we increase the number of observations, Gaussian matrix mean based
capacity estimation is seen to go very slowly towards the true capacity. To
illustrate this, figure~\ref{fig:L10n4r4} shows Gaussian matrix mean based capacity
estimation for $10$ observations on the same channel matrix.
It is seen that this decreases the deviation from the true capacity.

\section{Conclusion}
In this paper, we have shown that free probability provides a neat
framework for estimating the channel capacity for certain MIMO
systems. In the case of highly time varying environments, where one
can rely only on a set of limited noisy measurements, we have
provided an asymptotically unbiased estimator of the channel
capacity. A modified estimator called the Gaussian matrix mean based estimator 
was also introduced to take into account the bias in the case of finite
dimensions and was proved to be adequate for low rank channel
matrices. Moreover, although the results are based on asymptotic
claims (in the number of observations), simulations show that the
estimators work well for a very low number of observations also. Even
when considering discrepancies such as phase drifts and phase
off-set, the algorithm, based on the Gaussian matrix mean based estimator, provided
very good performance. Further research is being conducted to take
into account spatial correlation of the noise (in other words,
deconvolving with other measures than the  Mar\u{c}henko Pastur
law).

\appendices

\section{The proof of proposition~\ref{propfreeestimator}} \label{appendixb}
Let $(m_1,m_2,...)$ be the moments of $\eta$, $(M_1,M_2,...)$ the moments of $\eta\boxtimes\mu_c$. 
Then~\cite{eurecom:freedeconvinftheory}
\begin{equation} \label{momentcumulantc}
\begin{array}{lll}
  cM_1 &=& cm_1 \\
  cM_2 &=& cm_2 + c^2 m_1^2 \\
  cM_3 &=& cm_3 + 3 c^2 m_1 m_2 + c^3 m_1^3 \\
  cM_4 &=& cm_4 + 4 c^2 m_1 m_3 + 2 c^2 m_2^2 + 6 c^3 m_1^2 m_2 + c^4 m_1^4.
\end{array}
\end{equation}
Note that (\ref{momentcumulantc}) can also be inverted to express the $m_j$ in terms of the $M_j$ instead:
\begin{equation} \label{momentcumulantcreversed}
\begin{array}{l}
  cm_1=cM_1 \\
  cm_2=cM_2 - c^2 M_1^2\\
  cm_3=cM_3 - 3c^2 M_1 M_2 + 2c^3M_1^3\\
  cm_4=cM_4 - 4 c^2 M_1 M_3 - 2 c^2 M_2^2 + 10 c^3 M_1^2 M_2 - 5 c^4 M_1^4.
\end{array}
\end{equation}
Note also that the moments of $\eta\boxplus\delta_{\sigma^2}$ are
\begin{equation} \label{scalaradd}
\begin{array}{ll}
  &m_1 + {\sigma}^2 \\
  &m_2 + 2{\sigma}^2 m_1 + {\sigma}^4 \\
  &m_3 + 3 {\sigma}^2 m_2 + 3 {\sigma}^4 m_1 + {\sigma}^6 \\
  &m_4 + 4 \sigma^2 m_3 + 6 {\sigma}^4 m_2 + 4 {\sigma}^6 m_1 + {\sigma}^8
\end{array}
\end{equation}
By the definition of the free probability based estimator,
\[
  \nu_{ \frac{1}{m}\hat{\bf H}_{1...L}\hat{\bf H}_{1...L}^{H} }
  =
  \left( \left( \eta \boxslash \mu_{\frac{1}{L}} \right) \boxplus \delta_{\sigma^2}  \right)
  \boxtimes
  \mu_{\frac{1}{L}}
\]
where the moments of $\eta$ are $h_1,h_2,h_3,...$. Denoting by
$\eta_1 = \eta \boxslash \mu_{\frac{1}{L}}$, $\eta_2 = \eta_1
\boxplus \delta_{\sigma^2}$, we have that $\nu_{ \frac{1}{m}
\hat{\bf H}_{1...L}\hat{\bf H}_{1...L}^{H} } = \eta_2 \boxtimes
\mu_{\frac{1}{L}}$. Denote also the moments of $\eta_1$ by $r_i$,
the moments of $\eta_2$ by $s_i$, and as before the moments of
$\frac{1}{m} \hat{\bf H}_{1...L}\hat{\bf H}_{1...L}^{H}$ by
$\hat{h}_1,\hat{h}_2,\hat{h}_3,...$. Write also $c=\frac{n}{mL}$ as
in proposition~\ref{propfreeestimator}. For the third moment, we can
apply (\ref{momentcumulantc}), (\ref{scalaradd}) and
(\ref{momentcumulantcreversed}) in that order,
\[
\begin{array}{lll}
  \hat{h}_3 &=& s_3 + 3 c s_1 s_2 + c^2 s_1^3\\
            &=& r_3 + 3 {\sigma}^2 r_2 + 3 {\sigma}^4 r_1 + {\sigma}^6 \\
            & & + 3 c (r_1 + {\sigma}^2)(r_2 + 2{\sigma}^2 r_1 + {\sigma}^4) \\
            & & + c^2 (r_1 + {\sigma}^2)^3 \\
            &=& r_3 + 3 c r_1 r_2 + c^2 r_1^3 \\
            & & + 3 {\sigma}^2 (1+c) r_2 + (6c+3c^2) {\sigma}^2 r_1^2 \\
            & & + {\sigma}^4 (3+9c+3c^2) r_1 + {\sigma}^6 (1+3c+c^2) \\
            &=& h_3 - 3c h_1 h_2 + 2c^2h_1^3 \\
            & & + 3 c h_1 (h_2 - c h_1^2) + c^2 h_1^3 \\
            & & + (6c+3c^2) {\sigma}^2 h_1^2 + 3 {\sigma}^2 (1+c) (h_2 - c h_1^2) \\
            & & + {\sigma}^4 (3+9c+3c^2) h_1 + {\sigma}^6 (1+3c+c^2) \\
            &=& h_3 + 3{\sigma}^2(1+c) h_2 + 3{\sigma}^2 c h_1^2 \\
            & & + 3{\sigma}^4 \left( c^2+3c+1 \right) h_1 + {\sigma}^6 \left( c^2+3c+1 \right),
\end{array}
\]
which is the third equation in (\ref{freeestimator}) of proposition~\ref{propfreeestimator}.
Calculations are similar for the other moments, but more tedious for the fourth moment.

\section{The proof of proposition~\ref{propgaussian}} \label{appendixc0}
In all the following, the matrices are of dimension $n \times N$. We
need some terminology and results
from~\cite{paper:haagerupthorbjornsen1} for the proof of
proposition~\ref{propgaussian}. Let $S_p$ be the set of permutations
of $p$ elements $\{ 1,2,...,p\}$. For $\pi\in S_p$, let also
$\hat{\pi}$ be the permutation in $S_{2p}$ defined by
\begin{equation} \label{equivalencedef1}
\begin{array}{rcll}
  \hat{\pi}(2j-1) &=& 2\pi^{-1}(j), & (j\in\{ 1,2,...,p\})\\
  \hat{\pi}(2j)   &=& 2\pi(j)-1,    & (j\in\{ 1,2,...,p\}),
\end{array}
\end{equation}
let ${\sim}_{\hat{\pi}}$ denote the equivalence relation on $\{ 1,...,2p \}$ generated by the expression
\begin{equation} \label{equivalencedef2}
  j {\sim}_{\hat{\pi}} \hat{\pi}(j)+1, \mbox{ (addition formed mod. $2p$)},
\end{equation}
and let $k(\hat{\pi})$ and $l(\hat{\pi})$ denote the number of equivalence
classes of ${\sim}_{\hat{\pi}}$ consisting of even numbers or odd numbers, respectively.
Corollary 1.12 in~\cite{paper:haagerupthorbjornsen1} (slightly rewritten) states that
\begin{equation} \label{fromhaagerup}
  E\left[ tr_{n}\left( \left( \frac{1}{N} {\bf X} {\bf X}^H \right)^p \right) \right] = \frac{1}{nN^p} \sum_{\pi\in S_p} N^{k(\hat{\pi})} n^{l(\hat{\pi})},
\end{equation}
(\ref{firstthreemoments}) can thus be proved by calculating all values of
$k(\hat{\pi})$ and $l(\hat{\pi})$ for $\pi$ in $S_1$, $S_2$, $S_3$ and $S_4$.
We prove here the case $p=3$, to get an idea on how the calculations are performed.
For the six permutations in $S_3$ we obtain the following numbers by using (\ref{equivalencedef1}) and (\ref{equivalencedef2}):
\[
\begin{array}{|l|l|l|l|} \hline
  \pi     & \mbox{Equivalence classes of } {\sim}_{\hat{\pi}}     & k(\hat{\pi}) & l(\hat{\pi}) \\ \hline
  (1,2,3) & \{ \{ 1,3,5 \} , \{ 2 \} , \{ 4 \} , \{ 6 \} \}       & 3            & 1 \\ \hline
  (1,3,2) & \{ \{ 1,3 \} , \{ 2 \} , \{ 4,6 \} , \{ 5 \} \}       & 2            & 2 \\ \hline
  (2,1,3) & \{ \{ 1,5 \} , \{ 2,4 \} , \{ 3 \} , \{ 6 \} \}       & 2            & 2 \\ \hline
  (2,3,1) & \{ \{ 1 \} , \{ 2,4,6 \} , \{ 3 \} , \{ 5 \} \}       & 1            & 3 \\ \hline
  (3,1,2) & \{ \{ 1,3,5 \} , \{ 2,4,6 \} \}                       & 1            & 1 \\ \hline
  (3,2,1) & \{ \{ 1 \} , \{ 2,6 \} , \{ 3,5 \} , \{ 4 \} \}       & 2            & 2 \\ \hline
\end{array}
\]
Here $\pi =(i,j,k)$ means that $\pi(1)=i,\pi(2)=j,\pi(3)=k$.
Putting the numbers into (\ref{fromhaagerup}) we get
\[
\begin{array}{l}
  E\left[ tr_{n}\left( \left( \frac{1}{N} {\bf X} {\bf X}^H \right)^p \right) \right] \\
  = \frac{1}{nN^3} \left( N^3 n + N^2 n^2 + N^2 n^2 + N n^3 + N n + N^2 n^2 \right) \\
  = 1 + 3 \frac{n}{N} + \frac{n^2}{N^2} + \frac{1}{N^2} = 1+ 3c + c^2 + \frac{1}{N^2},
\end{array}
\]
which is the third equation in (\ref{firstthreemoments}).
We skip the computations for the other equations in (\ref{firstthreemoments}), since they are very similar and quite tedious,
since $S_p$ has $p!$ elements.

\section{The proof of theorems~\ref{teofreeestimatormain} and~\ref{teounbiasedestimatormain}} \label{appendixc}
We will first show the following:

\begin{lemma} \label{lemgeneralized}
For systems of type (\ref{system1}), the following holds when ${\bf R}_n$ is deterministic:
\begin{equation} \label{generalized}
\begin{array}{lll}
  E\left[ tr_n \left( {\bf W}_n   \right) \right] &=& m_1 + \sigma^2\\
  E\left[ tr_n \left( {\bf W}_n^2 \right) \right] &=& m_2 + 2{\sigma}^2 (1+c) m_1 + {\sigma}^4 (1 + c)\\
  E\left[ tr_n \left( {\bf W}_n^3 \right) \right] &=& m_3 + 3{\sigma}^2(1+c) m_2 + 3{\sigma}^2 c m_1^2 \\
                                                  & & + 3{\sigma}^4 \left( c^2+3c+1+\frac{1}{N^2}\right) m_1\\
                                                  & & + {\sigma}^6 \left( c^2+3c+1+\frac{1}{N^2}\right) \\
  E\left[ tr_n \left( {\bf W}_n^4 \right) \right] &=& m_4 + 4{\sigma}^2(1+c) m_3 + 8{\sigma}^2 c m_2 m_1 \\
                                                  & & + {\sigma}^4 (6c^2+16c+6+\frac{16}{N^2}) m_2 \\
                                                  & & + 14{\sigma}^4 c(1+c) m_1^2 \\
                                                  & & + 4{\sigma}^6 (c^3+ 6c^2+6c+1+\frac{5(c+1)}{N^2}) m_1 \\
                                                  & & + {\sigma}^8 \left( c^3 + 6c^2 + 6c + 1 + \frac{5(c+1)}{N^2}\right),
\end{array}
\end{equation}
where $m_j = tr_n\left( \left( \frac{1}{N}{\bf R}_n{\bf R}_n^{H} \right)^j \right)$.
\end{lemma}

We remark that it is the assumption that ${\bf X}_n$ is Gaussian 
which makes the mixed moments $E\left[ tr_n \left( {\bf W}_n^j \right) \right]$ 
expressible in terms of the individual moments $m_j$. 
Without the Gaussian assumption, there is no reason why such a relationship should hold. 
Also, while our statements are made only for the four first moments, we remark that 
similar relationships can be written down for higher moments also, 
which deviate from corresponding free probability based estimates 
only in terms of the form $\frac{1}{N^{2k}}$ 
(that the deviation terms are on this form is actually a consequence of theorem 1.13 of~\cite{paper:haagerupthorbjornsen1}). 

Before we prove lemma~\ref{lemgeneralized}, let us explain how it proves theorems~\ref{teofreeestimatormain} and~\ref{teounbiasedestimatormain}: 
We substitute $mL$ for $N$ (i.e. $c=\frac{n}{mL}$) for the case of $L$ observations,
$m$ for $N$ (i.e. $c=\frac{n}{m}$) for the case of one observation,
and ${\bf H}_{1...L}$ for ${\bf R}_n$ in lemma~\ref{lemgeneralized}.
Since the first two equations in (\ref{generalized}) coincide with the corresponding first two formulas in (\ref{freeestimator}) and (\ref{unbiasedestimator}), 
we see that the free probability based and the Gaussian matrix mean based estimators coincide for the first two moments in the case of only one observation,
and that they are both unbiased for these two moments (regardless of which model is used).
This proves the third statement of theorem~\ref{teounbiasedestimatormain}, and the statements on $h_{f1}$ and $h_{f2}$ in theorem~\ref{teofreeestimatormain}.

The third and fourth formulas in (\ref{unbiasedestimator}) are seen to equal the third and fourth formulas in (\ref{generalized}),
which explains why the Gaussian matrix mean based estimator has no bias in the third and fourth moments, 
thereby proving the first statement of theorem~\ref{teounbiasedestimatormain} 
(model (\ref{model2}) is also addressed due to the relationship (\ref{usethis})). 
The bias in the free probability based estimator is easily found by noting that the only differences between the third formula in (\ref{freeestimator})
and the third formula in (\ref{generalized}) are the terms $\frac{3{\sigma}^4}{m^2L^2}m_1$ and $\frac{{\sigma}^6}{m^2L^2}$.
This proves statements 2 in theorem~\ref{teofreeestimatormain}.

To see that $C_G$ is asymptotically unbiased when $m\rightarrow\infty$ (with $n,L$ kept fixed), 
it is sufficient to prove that the variance of all moments $tr_n({\bf W}_n^k)$ go to zero. 
This will remedy the fact that the capacity is a non-linear expression of the moments.
The proof for this part is a bit sketchy, since a similar analysis of such variances has already been done more throughly in connection with 
the theory of {\em second order freeness}~\cite{secondorderfreeness3}.
We need to analyse 
\begin{equation} \label{varianceequation}
  E\left( \left(tr_n({\bf W}_n^k)\right)^2\right) - \left( E(tr_n({\bf W}_n^k)) \right)^2.
\end{equation}
This analysis is very similar to the one in the proof of lemma~\ref{lemgeneralized} below: 
One simply associates each term in ${\bf W}_n^k$ with a circle with $2k$ edges, and identify the edges 
which correspond to equal, Gaussian elements (this corresponds to the equivalence relation $\sim_{\hat{\pi}}$ of appendix~\ref{appendixc0}). 
Computation of $E\left( \left(tr_n({\bf W}_n^k)\right)^2\right)$ and $\left( E(tr_n({\bf W}_n^k)) \right)^2$ is thus reduced to 
counting the number of terms which give rise to the different identifications of the edges on two circles (one circle for each trace).
We need only consider identifications which are pairings, 
due to the statements in appendix~\ref{appendixc0} when the matrix entries are Gaussian (see also~\cite{paper:thorbjornsen1,paper:haagerupthorbjornsen1}). 

One sees immediately that the edge identifications which can be found in $\left( E(tr_n({\bf W}_n^k)) \right)^2$ is a subset of the 
edge identifications which can be found in $E\left( \left(tr_n({\bf W}_n^k)\right)^2\right)$. 
These edge identifications therefore cancel each other in the expression for the variance, and we may therefore restrict to edge identifications 
which only appear in $E\left( \left(tr_n({\bf W}_n^k)\right)^2\right)$. 
These correspond to the edge identifications where at least one identification across the two circles takes place. 
If we perform one such edge identification first, we are left with one circle with $4k-2$ edges (when the two identified edges are skipped). 
After the identification of the remaining edges, the vertices can be associated with a choice among the elements $\{ 1,...,N\}$, or a choice among the elements 
$\{ 1,...,n\}$ (matching with matrix dimensions). 
Similarly as in appendix~\ref{appendixc0}, let $k(\hat{\pi})$ denote the number of vertices of the first type, 
$l(\hat{\pi})$ the number of vertices of the second type. 
It is clear that $k(\hat{\pi})\leq 2k-1$ after the identification of edges.
Since $N^{k(\hat{\pi})}\leq N^{2k-1}$ is not enough to cancel the leading $N^{2k}$-factor in $E\left( \left(tr_n({\bf W}_n^k)\right)^2\right)$ 
(recall that only $N$ goes to infinity, not $n$), we conclude that 
(\ref{varianceequation}) is $O\left(\frac{1}{N}\right)$, so that the variance of all moments go to $0$ as claimed, 
and we have established the second statement of theorem~\ref{teounbiasedestimatormain}.

$C_f$ is, following the same reasoning, asymptotically unbiased when $L\rightarrow\infty$ or $m\rightarrow\infty$ for model (\ref{model1}), and
when $L=1$ and $m\rightarrow\infty$ for model ({\ref{model2}). This proves the two second statements in theorem~\ref{teofreeestimatormain},
which concludes the proof of theorems~\ref{teofreeestimatormain} and~\ref{teounbiasedestimatormain}.

{\bf Proof of lemma~\ref{lemgeneralized}:}
Two facts are important in the proof. 
First of all, if $x_1,...,x_k$ are standard i.i.d. complex Gaussian random variables, then, according to remark 2.2 in~\cite{paper:thorbjornsen1},
\begin{equation} \label{simpleform}
   E(\left( x_1^{i_1} (\overline{x_1})^{j_1} \cdots x_k^{i_k} (\overline{x_k})^{j_k} \right) = 0 \mbox{ unless } i_1=j_1,...,i_k=j_k.
\end{equation}
Secondly, $E(|x_i|^{2p}) = p!$ for such $x_1,...,x_k$. 
we remark that the proof presented here can be simplified by using the following trick, taken from~\cite{paper:haagerupthorbjornsen1}: 
Rewrite a complex standard Gaussian random variable $x$ to the form 
$\frac{1}{\sqrt{s}} \left( x_1 + \cdots + x_s \right)$, 
where $x_1,...x_s$ are i.i.d. complex, standard, and Gaussian.  
(\cite{paper:haagerupthorbjornsen1} uses this trick, and lets $s$ go to infinity). 

Set ${\bf \Gamma}_n=\frac{1}{N}{\bf R}_n{\bf R}_n^{H}$.
Let us first look at the case for the second moment. Note that
\begin{equation} \label{tocompute2}
\begin{array}{lll}
  \lefteqn{E\left[ tr_n\left( {\bf W}_n^2 \right) \right]} \\
  &=& E\left[ tr_n\left( {\bf \Gamma}_n^2 \right) \right] \\
  & & + E\left[ tr_n\left( \sigma^2 \frac{1}{N^2} {\bf R}_n {\bf R}_n^H {\bf X}_n {\bf X}_n^H \right) \right] \\
  & & + E\left[ tr_n\left( \sigma^2 \frac{1}{N^2} {\bf R}_n {\bf X}_n^H {\bf X}_n {\bf R}_n^H \right) \right] \\
  & & + E\left[ tr_n\left( \sigma^2 \frac{1}{N^2} {\bf X}_n {\bf X}_n^H {\bf R}_n {\bf R}_n^H \right) \right] \\
  & & + E\left[ tr_n\left( \sigma^2 \frac{1}{N^2} {\bf X}_n {\bf R}_n^H {\bf R}_n {\bf X}_n^H \right) \right] \\
  & & + E\left[ tr_n\left( \sigma^4 \left( \frac{1}{N} {\bf X}_n {\bf X}_n^H \right)^2 \right) \right] + E\left[ tr_n\left( n_2 \right) \right],
\end{array}
\end{equation}
where the terms in $n_2$ have expectation zero due to (\ref{simpleform}). 
We see that
\begin{itemize}
  \item the first (deterministic) term is $m_2$,
    matching the first term in the second equation of (\ref{generalized}),
  \item The next-to-last term is $\sigma^4 (1+c)$, according to the second equation in (\ref{firstthreemoments}).
    This matches the last term in the second equation of (\ref{generalized}).
  \item By direct computation, the second term is
    \[ \sigma^2 \frac{1}{N^2n} \sum_{i,j,k,l} E\left( {\bf R}_n(i,j) {\bf R}_n^H (j,k) {\bf X}_n(k,l) {\bf X}_n^H (l,i). \right) \]
    This is nonzero only for $k=i$, so that this equals
    \[
    \begin{array}{l}
      \sigma^2 \frac{1}{N^2n} \sum_{i,j} N {\bf R}_n(i,j) {\bf R}_n^H (j,i) \\
      = \sigma^2 \frac{1}{N^2n} N n tr_n \left( {\bf R}_n {\bf R}_n^H \right) \\
      = \sigma^2 tr_n \left( \frac{1}{N} {\bf R}_n {\bf R}_n^H \right) = \sigma^2 m_1.
    \end{array}
    \]
  \item Similarly for the third term, which equals
    \[
    \begin{array}{l}
      \sigma^2 \frac{1}{N^2n} \sum_{i,j,k,l} E\left( {\bf R}_n(i,j) {\bf X}_n^H (j,k) {\bf X}(k,l) {\bf R}_n^H (l,i) \right) \\
      = \sigma^2 \frac{1}{N^2n} n n tr_n \left( {\bf R}_n {\bf R}_n^H \right) = \sigma^2 c m_1
    \end{array}
    \]
  \item The fourth and fifth term equal the second and third due to the trace property,
    so that the sum of the contributions of the second to fifth terms are $2 \sigma^2 (1+c) m_1$,
    which matches the second term in the second equation of (\ref{generalized}).
\end{itemize}
Thus, contributions on the right hand side of (\ref{tocompute2}) add up to the right hand side of the second equation in (\ref{generalized}), 
proving the case for the second moment.

For the third moment, write
\begin{equation} \label{tocompute3}
\begin{array}{l}
  E\left[ tr_n\left( {\bf W}_n^3 \right) \right] = E\left[ tr_n\left( {\bf \Gamma}_n^3 \right) \right] + \sigma^2 E\left[ tr_n\left( (\alpha_{31} + \alpha_{32}) \right) \right] \\
  + \sigma^4 E\left[ tr_n\left( (\beta_{31} + \beta_{32}) \right) \right] + \sigma^6 E\left[ tr_n\left( \left( \frac{1}{N} {\bf X}_n {\bf X}_n^H \right)^3 \right) \right] \\
  + E\left[ tr_n\left( n_3 \right) \right]
\end{array}
\end{equation}
where the terms in $n_3$ all have expectation zero, and
\[
\begin{array}{l}
  \alpha_{31} = \frac{1}{N^3} ( {\bf X}_n {\bf X}_n^H {\bf R}_n {\bf R}_n^H {\bf R}_n {\bf R}_n^H + {\bf R}_n {\bf X}_n^H {\bf X}_n {\bf R}_n^H {\bf R}_n {\bf R}_n^H \\
  + {\bf R}_n {\bf R}_n^H {\bf X}_n {\bf X}_n^H {\bf R}_n {\bf R}_n^H + {\bf R}_n {\bf R}_n^H {\bf R}_n {\bf X}_n^H {\bf X}_n {\bf R}_n^H \\
  + {\bf R}_n {\bf R}_n^H {\bf R}_n {\bf R}_n^H {\bf X}_n {\bf X}_n^H + {\bf X}_n {\bf R}_n^H {\bf R}_n {\bf R}_n^H {\bf R}_n {\bf X}_n^H),
\end{array}
\]
\[
\begin{array}{l}
  \alpha_{32} = \frac{1}{N^3} ( {\bf X}_n {\bf R}_n^H {\bf R}_n {\bf X}_n^H {\bf R}_n {\bf R}_n^H + {\bf R}_n {\bf X}_n^H {\bf R}_n {\bf R}_n^H {\bf X}_n {\bf R}_n^H \\
  + {\bf R}_n {\bf R}_n^H {\bf X}_n {\bf R}_n^H {\bf R}_n {\bf X}_n^H),
\end{array}
\]
\[
\begin{array}{l}
  \beta_{31} = \frac{1}{N^3}  ( {\bf R}_n {\bf R}_n^H {\bf X}_n {\bf X}_n^H {\bf X}_n {\bf X}_n^H + {\bf X}_n {\bf R}_n^H {\bf R}_n {\bf X}_n^H {\bf X}_n {\bf X}_n^H \\
  + {\bf X}_n {\bf X}_n^H {\bf R}_n {\bf R}_n^H {\bf X}_n {\bf X}_n^H + {\bf X}_n {\bf X}_n^H {\bf X}_n {\bf R}_n^H {\bf R}_n {\bf X}_n^H \\
  + {\bf X}_n {\bf X}_n^H {\bf X}_n {\bf X}_n^H {\bf R}_n {\bf R}_n^H + {\bf R}_n {\bf X}_n^H {\bf X}_n {\bf X}_n^H {\bf X}_n {\bf R}_n^H),
\end{array}
\]
\[
\begin{array}{l}
  \beta_{32} = \frac{1}{N^3}  ( {\bf R}_n {\bf X}_n^H {\bf X}_n {\bf R}_n^H {\bf X}_n {\bf X}_n^H + {\bf X}_n {\bf R}_n^H {\bf X}_n {\bf X}_n^H {\bf R}_n {\bf X}_n^H \\
  + {\bf X}_n {\bf X}_n^H {\bf R}_n {\bf X}_n^H {\bf X}_n {\bf R}_n^H)
\end{array}
\]
(i.e. the terms in $\alpha_{31},\beta_{31}$ have the terms ${\bf X}_n$,${\bf X}_n^H$ adjacent to each other). 
We see that
\begin{itemize}
  \item the first and fourth terms in (\ref{tocompute3}) match the first and fifth terms on the right hand side of the third equation in (\ref{generalized})
    (due to (\ref{firstthreemoments})).
  \item Three of the terms in $\alpha_{31}$ are seen to contribute with
    \[
      \frac{1}{N^3n} N n tr_n \left( \left( {\bf R}_n {\bf R}_n^H \right)^2 \right) = m_2,
    \]
    and the remaining three terms are seen to contribute
    \[
      \frac{1}{N^3n} n n tr_n \left( \left( {\bf R}_n {\bf R}_n^H \right)^2 \right) = c m_2
    \]
    Addition gives $\alpha_{31} = 3(1+c)m_2$.
  \item All terms in $\alpha_{32}$ are seen to contribute
    \[
      \frac{1}{N^3n} n n \left( tr_n \left( {\bf R}_n {\bf R}_n^H \right) \right)^2 = c m_1^2,
    \]
    so that the total contribution is $3cm_1^2$.
  \item Using the second formula in (\ref{firstthreemoments}), three terms in $\beta_{31}$ are seen to contribute
    \[
      \frac{1}{Nn} n tr_n \left( {\bf R}_n {\bf R}_n^H \right)  \frac{1}{n} n(1+c) = (1+c) m_1,
    \]
    and the remaining three terms contribute
    \[
      \frac{1}{Nn} n tr_n \left( {\bf R}_n {\bf R}_n^H \right)  \frac{1}{N} n(1+c) = c (1+c) m_1,
    \]
    Addition gives $3(c^2+2c+1)m_1$.
  \item All terms in $\beta_{32}$ are seen to contribute
    \[
      \frac{1}{N^3n} n tr_n \left( {\bf R}_n {\bf R}_n^H \right) (nN - 1) + \frac{1}{N^3n} n tr_n \left( {\bf R}_n {\bf R}_n^H \right) \times 2,
    \]
    where the factor $2$ comes in since $E(|x|^4) = 2$ for a complex standard Gaussian random variable.
    Simplifying we get $(c+\frac{1}{N^2})m_1$,
    so that the total contribution is $3(c+\frac{1}{N^2})m_1$
\end{itemize}
Thus, contributions on the right hand side of (\ref{tocompute3}) add up to the right hand side of the third equation in (\ref{generalized}), 
proving the case for the third moment also.

Now for the fourth equation in (\ref{generalized}). 
The details in this are similar to the
calculations for the third moment, but much more tedious. The first term for
the fourth moment formula in (\ref{generalized}) is trivial,
as is the last term which comes from the fourth formula in
(\ref{firstthreemoments}). The second and third terms are calculated
using exactly the same strategy as for the third moment. 
The remaining fourth, fifth and sixth terms require much
attention. We address just some of these. 

Computing $E\left[ tr_n\left( \sigma^6 (\beta_{41} + \beta_{42}) \right) \right]$ gives the sixth term, 
where the terms in $\beta_{41}$ are similar to those for $\beta_{31}$ (i.e. the terms ${\bf X}_n$,${\bf X}_n^H$ are adjacent to each other), 
i.e. four terms have the same trace as
\[
  a = \frac{1}{N^4} {\bf R}_n {\bf R}_n^H {\bf X}_n {\bf X}_n^H {\bf X}_n {\bf X}_n^H {\bf X}_n {\bf X}_n^H,
\]
while four terms have the same trace as 
\[
  b = \frac{1}{N^4} {\bf X}_n {\bf R}_n^H {\bf R}_n {\bf X}_n^H {\bf X}_n {\bf X}_n^H {\bf X}_n {\bf X}_n^H,
\]
It is clear that $E\left[ tr_n(a) \right]$ equals
\begin{eqnarray*}
  & & \frac{1}{N^4n} n tr_n \left( {\bf R}_n {\bf R}_n^H \right) E\left[ tr_n\left( \left( {\bf X}_n {\bf X}_n^H \right)^2 \right) \right] \\
  &=& tr_n \left( \frac{1}{N} {\bf R}_n {\bf R}_n^H \right) E\left[ tr_n\left( \left( \frac{1}{N} {\bf X}_n {\bf X}_n^H \right)^2 \right) \right] \\
  &=& \left( c^2+3c+1+\frac{1}{N^2}\right) m_1, 
\end{eqnarray*}
and that $E\left[ tr_n(b) \right]$ equals
\begin{eqnarray*}
  & & \frac{1}{N^4n} n tr_n \left( {\bf R}_n {\bf R}_n^H \right) c E\left[ tr_n\left( \left( {\bf X}_n {\bf X}_n^H \right)^2 \right) \right] \\
  &=& c tr_n \left( \frac{1}{N} {\bf R}_n {\bf R}_n^H \right) E\left[ tr_n\left( \left( \frac{1}{N} {\bf X}_n {\bf X}_n^H \right)^2 \right) \right] \\
  &=& c\left( c^2+3c+1+\frac{1}{N^2} \right) m_1, 
\end{eqnarray*}
so that $\beta_{41}=4(1+c)\left( c^2+3c+1+\frac{1}{N^2}\right) m_1$. 

Similarly, for $\beta_{42}$ (where the terms ${\bf X}_n$,${\bf X}_n^H$ are not adjacent to each other), 
we need to address four terms which all have the same trace as
\[
  c = \frac{1}{N^4} {\bf R}_n {\bf X}_n^H {\bf X}_n {\bf R}_n^H {\bf X}_n {\bf X}_n^H {\bf X}_n {\bf X}_n^H,
\]
and four terms which have the same trace as 
\[
  d = \frac{1}{N^4} {\bf X}_n {\bf R}_n^H {\bf X}_n {\bf X}_n^H {\bf R}_n {\bf X}_n^H {\bf X}_n {\bf X}_n^H.
\]
By counting terms carefully, we see that these eight terms together contribute with $\left( 8c+8c^2+\frac{16(c+1)}{N^2} \right) m_1$ 
(during this count of terms, we need the fact  that $E(|x|^6) = 6$ when $x$ is complex, standard, and Gaussian). 
All in all we have that
\begin{eqnarray*}
  \lefteqn{E\left[ tr_n\left( \sigma^6 (\beta_{41} + \beta_{42}) \right) \right] = } \\
  & & 4\sigma^6 (1+c)\left( c^2+3c+1+\frac{1}{N^2}\right) m_1 + \\
  & & \sigma^6 \left(8c+8c^2+\frac{16(c+1)}{N^2} \right) m_1 \\
  &=& 4\sigma^6 \left( c^3+6c^2+6c+1+\frac{5(1+c)}{N^2} \right) m_1,
\end{eqnarray*}
which is the sixth term in the fourth equation of (\ref{generalized}). 

The details for the fourth and fifth terms are dropped. 
\sluttmerke

As can be seen, the requirement that ${\bf R}_n$ is deterministic is not strictly necessary in the proof of lemma~\ref{lemgeneralized}, 
so that we could replace it with any random matrix independent from ${\bf X}_n$, 
the moment $m_j$ with $E\left[ tr_n\left( \left( \frac{1}{N}{\bf R}_n{\bf R}_n^{H} \right)^j \right) \right]$, 
and $m_j^2$ with $E\left[ \left( tr_n\left( \left( \frac{1}{N}{\bf R}_n{\bf R}_n^{H} \right)^j \right) \right)^2 \right]$.

\bibliography{../bib/BSTLabbrev,../bib/mybib,../bib/mainbib}

\end{document}